\documentclass[superscriptaddress,showpacs,amssymb,11pt,reprint,aps,prd,longbibliography,nofootinbib,floatfix]{revtex4-1}

\usepackage{graphicx,epsfig,amssymb,times} 
\usepackage{amsmath,amsfonts}
\usepackage{bm}
\usepackage{epstopdf}
\usepackage{hyperref}
\usepackage[caption=false]{subfig}
\usepackage[usenames]{color}   
\usepackage[dvipsnames]{xcolor}

\usepackage[normalem]{ulem}

\definecolor{coolblack}{rgb}{0.0, 0.18, 0.39}
\definecolor{darkred}{rgb}{0.5,0,0}
\definecolor{darkgreen}{rgb}{0,0.5,0}
\definecolor{darkblue}{rgb}{0,0,0.5}
\definecolor{lapislazuli}{rgb}{0.15, 0.38, 0.61}
\definecolor{venetianred}{rgb}{0.78, 0.03, 0.08}
\definecolor{bleudefrance}{rgb}{0.19, 0.55, 0.91}
\definecolor{dogwoodrose}{rgb}{0.84, 0.09, 0.41}
\hypersetup{colorlinks=true, citecolor=darkblue, linkcolor=darkblue, 
urlcolor = darkblue}

\newcommand{\beq}{\begin{equation}}
\newcommand{\eeq}{\end{equation}}

\newcommand{\newtext}[1]{#1}

\begin{document}

\title{\large Superradiant instability of a charged regular black hole}
	
	\author{Sam R. Dolan}
	\email{s.dolan@sheffield.ac.uk}
	\affiliation{Consortium for Fundamental Physics, School of Mathematics and Statistics, University of Sheffield, Hicks Building, Hounsfield Road, Sheffield S3 7RH, United Kingdom.}
	
	\author{Marco A. A. de Paula}
	\email{marco.paula@icen.ufpa.br}
	\affiliation{Programa de P\'os-Gradua\c{c}\~{a}o em F\'{\i}sica, Universidade 
		Federal do Par\'a, 66075-110, Bel\'em, Par\'a, Brazil.}
	
	\author{Luiz C. S. Leite}
	\email{luiz.leite@ifpa.edu.br}
	\affiliation{Campus Altamira, Instituto Federal do Par\'a, 68377-630, Altamira, Par\'a, Brazil.}
	
	\author{Lu\'is C. B. Crispino}
	\email{crispino@ufpa.br}
	\affiliation{Programa de P\'os-Gradua\c{c}\~{a}o em F\'{\i}sica, Universidade 
		Federal do Par\'a, 66075-110, Bel\'em, Par\'a, Brazil.}

\begin{abstract}
We show that a charged, massive scalar field in the vicinity of an electrically-charged Ay\'on-Beato-Garc\'ia (ABG) regular black hole has a spectrum of quasibound states that (in a certain parameter regime) grow exponentially with time, due to black hole superradiance. Superradiant quasibound states are made possible by the enhancement of the electrostatic potential at the horizon in nonlinear electrodynamics; in contrast, the Reissner-Nordstr\"om black hole does not appear to possess such superradiant quasibound  states. Here we compute the spectrum for a range of multipoles $\ell$ across the parameter space, and we find the fastest growth rate in the monopole mode. We find that a regular black hole with a small charge can still trigger a significant superradiant instability if the charge-to-mass ratio of the field is compensatingly large. \newtext{We estimate the amount of black hole mass that can be deposited in the scalar field, finding an upper bound of circa $20\%$ in the extreme charge scenario.} Finally, we consider the stationary bound states at the superradiant threshold, and we conjecture that, due to this instability, the ABG black hole will evolve towards a configuration with charged scalar hair.
\end{abstract}

\date{\today}

\maketitle


\section{Introduction}

The stability of black holes under perturbation has been scrutinized for at least six decades \cite{Regge:1957td}. Questions of stability are rendered more subtle by the phenomenon of superradiance \cite{Brito:2015oca}, in which an impinging wave is \emph{amplified} as it scatters from a black hole \cite{zel1972amplification, Misner:1972, Starobinsky:1973aij, Starobinskil:1974nkd}. If superradiant flux is subsequently reflected back onto the hole, then an instability may develop. In the illustrative `black hole bomb' proposal \cite{Press:1972zz}, reflection is achieved by means of a surrounding mirror. However, reflection and confinement may arise more naturally in certain scenarios; for example, if the spacetime has a boundary \cite{Cardoso:2004hs, Cardoso:2006wa}, or (it has been argued) if the black hole is surrounded by a plasma~\cite{Pani:2013hpa,Conlon:2017hhi,  Cannizzaro:2020uap,Cannizzaro:2021zbp, Dima:2020rzg}, a torus magnetosphere \cite{VanPutten:1999vda}, or an accretion disk \cite{Lingetti:2022psy}. In particular, if the perturbing field has a mass, then (due to the long-ranged Newtonian attraction and centrifugal repulsion) the potential will have a minimum, and the field will possess a spectrum of quasibound states, confined within the vicinity of the black hole \cite{Ternov:1980st, Gaina:1992nx, Lasenby:2002mc, Baumann:2019eav}. If these states are also within the superradiant regime, then they will grow exponentially with time, and an instability ensues~\cite{Damour:1976kh, Detweiler:1980uk, Zouros:1979iw, Cardoso:2005vk, Dolan:2007mj, Arvanitaki:2009fg,Arvanitaki:2010sy,Witek:2012tr, Dolan:2012yt, Brito:2014wla, Yoshino:2015nsa, East:2017ovw, Frolov:2018ezx, Dias:2023ynv}. \newtext{See also Refs.~\cite{PhysRevD.91.124026,PhysRevD.99.104039,PhysRevD.101.044055,PhysRevD.103.084019,PhysRevD.105.064043} and references therein.}

Superradiance can be understood as a consequence of the laws of black hole mechanics \cite{Bardeen:1973gs}. The first law states that
\beq
d M = \frac{\kappa_H}{8 \pi} dA + \Omega_H d J + \Phi_H d Q , 
\eeq
where $M$, $A$, $J$ and $Q$ are the mass, area, angular momentum and charge of the hole; and $\kappa_H$, $\Omega_H$ and $\Phi_H$ are its surface gravity, angular frequency and electrostatic potential at the (outer) horizon. The second law states that the area of the event horizon of a black hole does not decrease with time in a classical process: $dA \ge 0$. Hence, in a process that increases the horizon area (and thus the black hole entropy), 
\beq
\frac{8\pi}{\kappa_H} \left( 1 - \Omega_H \frac{dJ}{dM} - \Phi_H \frac{dQ}{dM} \right) dM > 0 .
\eeq
If the quantity in parantheses is negative (and $\kappa_H$ remains positive), then $dM < 0$, that is, the black hole loses mass into the field. For a (mode of a) field of frequency $\omega > 0$, charge $q$ and azimuthal angular momentum number $m$, one can replace $dJ/dM = m / \omega$ and $dQ/dM = q / \omega$. Hence the condition for superradiance is
\beq
0 < \omega < \omega_c , \quad \quad \omega_c \equiv m \Omega_H + q \Phi_H . \label{eq:superradiance-regime}
\eeq
In this superradiant regime the black hole transfers mass and angular momentum (and/or charge) into the perturbing field.

Astrophysical black holes likely possess significant angular momentum but negligible charge \cite{Gibbons:1975kk}. (That said, a tiny amount of charge can significantly affect the dynamics of charged particles in the black hole vicinity~\cite{Schroven:2017jsp,Zajacek:2018ycb,Zajacek:2018vsj}.) The majority of studies in the literature focus on \emph{rotational} superradiance. For Kerr (and Kerr-Newman) black holes the spectrum of quasibound states is well-characterised for massive scalar fields \cite{Zouros:1979iw,Detweiler:1980uk,Furuhashi:2004jk,Cardoso:2005vk,Dolan:2007mj,Dolan:2012yt}, Proca fields \cite{Pani:2012bp,Pani:2012vp,Baryakhtar:2017ngi,East:2017ovw,East:2017mrj,Cardoso:2018tly,Frolov:2018ezx,Dolan:2018dqv}, and massive spin-two fields \cite{Brito:2020lup,Dias:2023ynv}. Quasibound states lie within the superradiant regime if the mass of the field is sufficiently small, $M \mu \sim 1/2$, where 
\beq
M \mu \equiv \dfrac{G M \mu}{\hbar c} \approx  7.5 \times 10^{9} \left(\dfrac{M}{M_{\odot}}\right) \left(\dfrac{\mu}{1 \text{eV}} \right)
\eeq
plays the role of an effective gravitational fine-structure constant. In other words, for a significant instability the Compton wavelength of the field must be comparable with the gravitational radius of the black hole. Hence the superradiant instability is negligible for Standard Model (SM) fields perturbing astrophysical black holes, but it may be stimulated by ultralight fields such as axions~\cite{Arvanitaki:2009fg,Arvanitaki:2010sy} (or heavier fields coupled to primordial black holes \cite{Pani:2013hpa,Ferraz:2020zgi, Branco:2023frw, Calza:2023rjt}) so that black holes become detectors for ultralight beyond-SM particles. Attention in the last decade has moved on to characterising the observational signatures that may be generated \cite{Pani:2012vp, Cardoso:2018tly, Chia:2020dye, Chen:2022kzv, Siemonsen:2022ivj, Ghoshal:2023fno}, and quantifying how the instability evolves over long time periods and into the non-linear regime \cite{Witek:2012tr, Yoshino:2015nsa, Dolan:2012yt, Bamber:2020bpu}, as well as determining how much mass and angular momentum is ultimately transferred from the black hole into the field \cite{East:2017ovw, East:2017mrj}.

\begin{figure}[!htbp]
    \includegraphics[width=1.0\columnwidth]{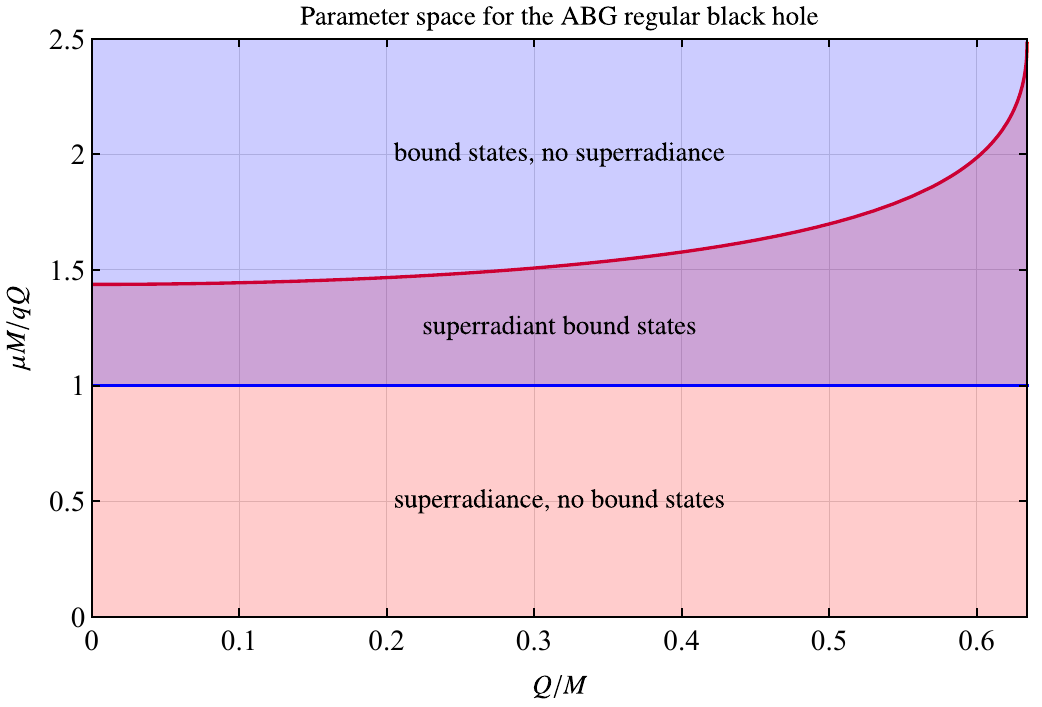}\vspace*{0.2cm}
    \includegraphics[width=1.0\columnwidth]{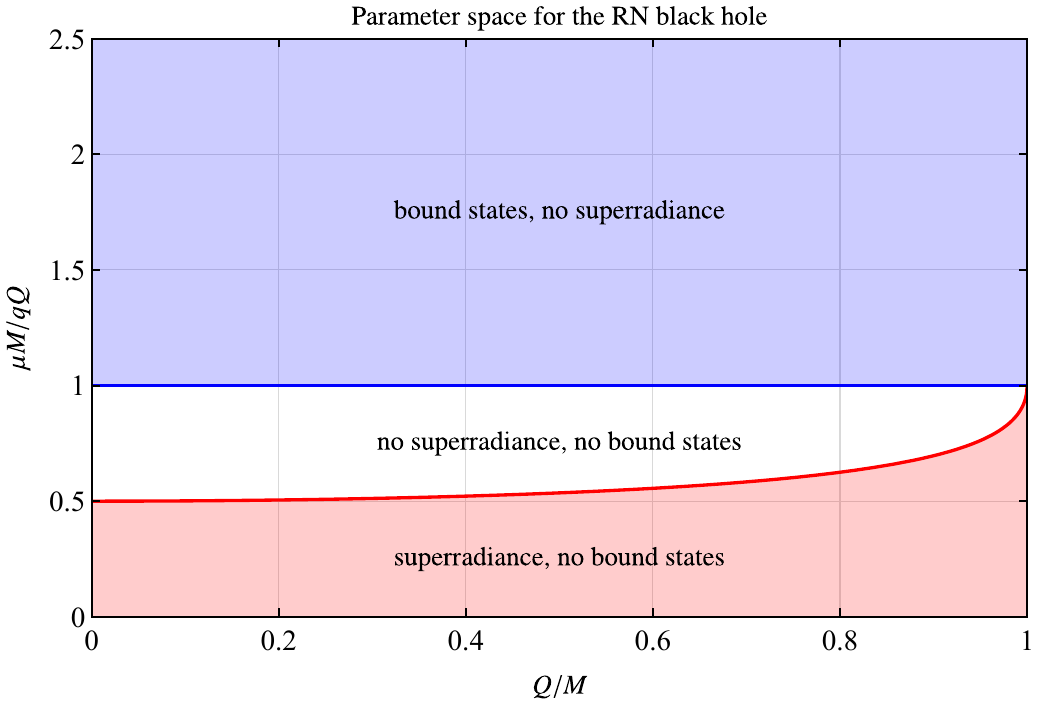} 
    \caption{
      Parameter space for the ABG black hole (upper) and the RN black hole (lower). For a RN black hole, the superradiant and bound-state regions are disjoint (though they meet at $Q=M$). For an ABG regular black hole, there is a wide region of overlap in which superradiant bound states exist. The superradiant-threshold line meets the vertical axis at $23/16$.
    }
    \label{fig:parameter-space}
\end{figure}

In principle, by Eq.~(\ref{eq:superradiance-regime}), instabilities driven by \emph{charged} superradiance are also possible. Unlike in the rotating case, a charged instability would arise principally in the spherically-symmetric (i.e.~$\ell = 0$) sector, making it far easier to track the development of the instability in the non-linear regime (i.e.~accounting for back-reaction in the Einstein-Maxwell equations). However,  studies have established that the simplest example of a charged black hole -- the Reissner-Nordstr\"om (RN) solution -- does \emph{not} appear to permit a superradiant instability~\cite{Hod:2012wmy,Hod:2015hza}, unless one confines the field in a cavity with a mirrorlike boundary condition, or a spacetime boundary~\cite{Herdeiro:2013pia, Dolan:2015dha, Dias:2018zjg,Davey:2021oye, Feiteira:2024awb}. Heuristically, the reason is clear: superradiant modes have a positive charge coupling $qQ$ that generates an electrostatic repulsion that is sufficient to overcome the gravitational attraction $\mu M$. This prevents bound states from forming in the  superradiant regime. Put another way, though superradiant modes exist and quasibound states exist, they do not exist in the same region of parameter space (see Fig.~\ref{fig:parameter-space}, lower plot). 

Recently, it has been shown that the situation is quite different for electrically-charged Ay\'on-Beato-Garc\'ia (ABG) regular black holes \cite{dePaula:2024xnd}. Regular black holes -- black holes seemingly free from curvature singularities -- arise in certain theories of \emph{nonlinear} electrodynamics \cite{Ayon-Beato:1998hmi,Ayon-Beato:1999kuh,Ayon-Beato:1999qin,Bronnikov:2000vy,dymnikova} (for a recent review see Ref.~\cite{Bronnikov:2022ofk}). \newtext{The ABG spacetime
was the first exact charged regular black hole solution obtained within the framework of general relativity minimally coupled to a nonlinear electrodynamics model \cite{Ayon-Beato:1998hmi}. In the exterior region, the ABG spacetime resembles the RN spacetime: it has the same weak-field limit as $r \rightarrow \infty$, as well as a similar causal structure featuring a light-ring and a non-degenerate event horizon. Thus, the ABG regular black hole provides a clean setting for exploring the imprints of nonlinear electrodynamics theory on black hole physics, and for drawing comparisons of the observable-in-principle properties of singular and regular black holes. }

In comparison with a RN black hole\newtext{, e.g.,} an ABG black hole has a stronger electrostatic potential at the horizon $\phi(r_{+})$ (for given $Q$ and $M$), and thus an enhanced superradiant regime. As shown in Fig.~\ref{fig:parameter-space} (see also Fig.~5 of Ref.~\cite{dePaula:2024xnd}), there is a region in parameter space where superradiance and quasibound states coexist, and this `overlap region' persists even in the weakly-charged limit ($Q/M \ll 1$). The main aim of this work is to compute the superradiant quasibound states associated with the overlap region of parameter space, and to characterize their spectrum and growth rates. 

The parameter space $(Q/M, \mu M/ q Q)$ shown in Fig.~\ref{fig:parameter-space} has been partitioned with two curves. Firstly, a hydrogenic spectrum of bound states will arise if the potential has an \emph{attractive} long-ranged $1/r$ term, which requires that $\alpha > 0$, where 
\beq
\alpha \equiv M \mu - Q q \label{eq:alpha}
\eeq
is the effective fine-structure constant (see Sec.~\ref{subsec:quasibound}). Secondly, from Eq.~(\ref{eq:superradiance-regime}), modes are superradiant if 
\beq
0 < \omega < \omega_c, \quad \quad \omega_c \equiv q \phi(r_+) , \label{eq:omegac}
\eeq
and for bound states $\omega \lesssim \mu$ (see Eq.~(\ref{eq:hydrogenic})). The red curve in Fig.~\ref{fig:parameter-space} denotes the approximate boundary of the superradiant region, determined by $\mu = \omega_c$. 

The remainder of this article is arranged as follows. In Sec.~\ref{sec:analysis}, we cover the ABG spacetime (\ref{subsec:spacetime}), superradiance in a massive scalar field (\ref{subsec:superradiance}), quasibound states (\ref{subsec:quasibound}), and the hydrogenic (\ref{subsec:hydrogenic}) and power-law approximations (\ref{subsec:growth}) for the spectrum and instability growth rate. The method for calculating the quasibound spectrum is outlined in Sec.~\ref{subsec:numerical-method}. A selection of numerical results is presented in Sec.~\ref{sec:results}. The paper concludes in Sec.~\ref{sec:conclusions} with a discussion that outlines some future work and a conjecture. Throughout this work we use the natural units, for which $G = c = \hbar = 4 \pi \epsilon_0 = 1$, and metric signature $+2$.


\section{Analysis\label{sec:analysis}}

\subsection{The spacetime\label{subsec:spacetime}}
The ABG spacetime is static and spherically symmetric. In polar coordinates $x^\mu = \{ t, r, \theta, \varphi \}$, the line element is
\beq
ds^2 = - f(r) dt^2 + f(r)^{-1} dr^2 + r^2 d\Omega^2 ,  \label{eq:le}
\eeq
where $d\Omega^2 = d\theta^2 + \sin^2 \theta d\varphi^2$, $f(r)$ is the metric function, and the electromagnetic vector potential is $A_{\mu}=(-\phi(r),0,0,0)$, with
\begin{align}
f(r) &= 1-\frac{2Mr^{2}}{(r^{2}+Q^{2})^{3/2}}+\frac{Q^{2}r^{2}}{(r^{2}+Q^{2})^{2}},
\label{eq:fABG}  \\
\phi(r) &= \frac{r^{5}}{2Q}\left(\dfrac{3M}{r^{5}}+\frac{2Q^{2}}{\left(Q^{2}+r^{2}\right)^{3}}-\frac{3M}{\left(Q^{2}+r^{2}\right)^{5/2}}\right). 
\label{eq:phiABG}
\end{align}

When the condition $|Q| \le Q_{\rm{ext}}\approx 0.6341M$ is fulfilled~\cite{Ayon-Beato:1998hmi}, the line element~\eqref{eq:le} describes an ABG RBH, with a Cauchy (inner) $r_{-}$ and an event (outer) $r_{+}$ horizon. For more details, see e.g.~Sec.~II of Refs.~\cite{dePaula:2024xnd,Paula:2020yfr}.

\subsection{The scalar field and superradiance\label{subsec:superradiance}}
The scalar field $\Phi$ satisfies the field equation
\begin{equation}
\label{KG}
\left(\nabla_{\nu} - iqA_{\nu}\right)\left(\nabla^{\nu} - iqA^{\nu}\right)\Phi - \mu^{2}\Phi = 0 ,
\end{equation}
where $\mu$ and $q$ correspond to the mass and charge of the field, respectively, and $\nabla_\mu$ denotes the covariant derivative. 
Exploiting the separability of Eq.~\eqref{KG}, we can write a particular mode of $\Phi$ as
\begin{equation}
\label{PHI}
\Phi\equiv \frac{1}{r} u_{\omega \ell}(r) Y_{\ell m} (\theta, \phi) e^{-i\omega t},
\end{equation}
in which $u_{\omega \ell}(r)$ is a radial function and $Y_{\ell m}$ are the spherical harmonics, with $\omega$ and $l$ being the frequency and angular momentum of the scalar field, respectively. We 
henceforth drop the subscripts from $u_{\omega \ell}(r)$ for brevity. After inserting Eq.~\eqref{PHI} into Eq.~\eqref{KG}, one obtains a radial equation of the form
\begin{equation}
\left\{ f(r) \frac{d}{dr}\left[ f(r) \frac{d}{d r}\right] - V(r) \right\} u(r) = 0, 
\label{eq:radial} 
\end{equation}
where $f(r)$ is defined in Eq.~(\ref{eq:fABG}), and the potential function $V(r)$ reads
\begin{equation}
\label{EffP} V(r) \equiv f(r)\left[\mu^{2}+\frac{1}{r}\dfrac{df(r)}{dr}+ \dfrac{l(l+1)}{r^{2}}\right]-\left[\omega-q \phi(r)\right]^{2}.
\end{equation}
The radial function $u(r)$ is evanescent (i.e.,~exponential) in regions where $V(r) > 0$, and propagative (i.e.,~oscillatory) in regions where $V(r) < 0$.

At the outer horizon, $f(r_+) = 0$ and $V(r_+) = - \tilde{\omega}^2$, where 
\beq
\tilde{\omega} \equiv \omega - \omega_c,
\eeq
and $\omega_c$ is defined in Eq.~(\ref{eq:omegac}). In the region $|r - r_+| \ll M$, the radial equation takes the form
\beq
\left\{ \frac{d^2}{d r_\star^2} + \tilde{\omega}^2 \right\} u (r) = 0 , 
\eeq
where $r_{\star}$ is the \emph{tortoise coordinate} defined by $dr_{\star} \equiv dr/f(r)$. In general, $u(r)$ is a linear sum of $e^{i \tilde{\omega} r_\star}$ and $e^{-i \tilde{\omega} r_\star}$. In Sec.~IIIA of Ref.~\cite{dePaula:2024xnd} it is established that only the latter term is regular on the future horizon in a horizon-penetrating coordinate system.
Further, by standard arguments it is then shown that the transmission factor exceeds unity (i.e.~the mode is superradiant) if $\tilde{\omega} < 0$, which is in accordance with Eq.~(\ref{eq:omegac}). Relatedly, in the superradiant regime there is a flow of charge \emph{out} of the horizon, as one can see by contracting the charge current
\beq
J_\mu = \frac{q}{2 i} \left( \Phi^\ast (\nabla_\mu - i q A_\mu) \Phi - \Phi \left( \nabla_\mu + i q A_\mu \right) \Phi^\ast \right) 
\eeq
with the normal to a constant-$r$ hypersurface evaluated at the horizon  $n^\mu = g^{\mu \nu} \nabla_\nu r$, to get
\begin{align}
\left. n^\mu J_\mu \right|_{r_+} &= \frac{q}{2 i} \left( u^\ast \frac{d u}{d r_\star} - u \frac{d u^\ast}{d r_\star} \right) \left|Y_{\ell m} \right|^2 
= - q \tilde{\omega}  \left|Y_{\ell m} \right|^2 .
\end{align}
The sign of the flux depends on the sign of $\tilde{\omega}$. 

In the far-field, at leading order, Eq.~(\ref{eq:radial}) reduces to 
\beq
\left\{ \dfrac{d^2}{d r_\star^2} - k^2 \right\} u (r) = 0,
\eeq
with $k^2 \equiv \mu^2 - \omega^2$ and solutions $e^{\pm k r_\ast}$.

\subsection{Quasibound states\label{subsec:quasibound}}
Here, we are interested in solutions which are regular on the future horizon in a horizon-penetrating coordinate system, and which decay in the far field.
More precisely, we seek quasibound state solutions satisfying the following boundary conditions
\begin{equation}
\label{BC}\Psi_{\omega l}\sim\begin{cases}
e^{-i \tilde{\omega} r_{\star}}, & r_{\star}\rightarrow -\infty , \\
R_{\omega l}e^{- k r_\star}, & r_\star \rightarrow+\infty ,
\end{cases}
\end{equation}
where $\tilde{\omega} \equiv \omega - q \phi_+$ (with $\phi_+ \equiv \phi(r_+)$), $k \equiv \sqrt{\mu^{2}-\omega^{2}}$, and $R_{\omega l}$ are complex coefficients. These boundary conditions are only satisfied at the distinct (in general, complex) values of $\omega_{\ell n}$ in the quasibound state spectrum.

\subsubsection{The hydrogenic approximation\label{subsec:hydrogenic}}
We now consider the form of the spectrum when $M \mu \ll 1$ and $q Q \ll 1$. In this regime, the bound states extend to large radii, and a robust understanding of the real part of the spectrum can be gained by expanding the radial equation at large $r$ \cite{Detweiler:1980uk}. Expanding the potential in inverse powers of $1/r$, Eq.~(\ref{eq:radial}) reduces to
\beq
\left\{ \frac{d^2}{dr^2} - k^2 + \frac{2 \nu k}{r} - \frac{\ell (\ell + 1) + \beta}{r^2} \right\} u(r) = 0 , \label{eq:farfield}
\eeq
where $k^2 \equiv \mu^2 - \omega^2$, $\nu k \equiv M \mu^2 - q Q \omega$ and $\beta$ is small for $M\mu \ll 1$ and neglected in this section (see Ref.~\cite{Bao:2022hew} for subtleties when calculating the growth rate). For bound states to emerge we require that $k^2 > 0$ and $\nu k > 0$. 

The solution of Eq.~(\ref{eq:farfield}) with the correct boundary condition in the far-field is
\beq
u(r) = e^{- k r} (2 k r)^{\ell + 1} U(\ell + 1 - \nu, 2 \ell + 2, 2 k r) .  \label{eq:hydrogenic-wf}
\eeq
The bound states corresponds to integer values of $\nu$ such that $\nu = \ell + 1 + n \equiv \bar{n}$, for which the confluent hypergeometric function  $U(-n, 2 \ell + 2, 2 k r)$ reduces to a Laguerre polynomial. Here $n = 0, 1, \ldots$ is the excitation number and $\bar{n} = \ell + 1, \ell + 2, \ldots$ is the principal quantum number. Hence, there emerges a discrete spectrum whose frequencies $\omega$ are solutions to 
\beq
\frac{M \mu^2 - q Q \omega}{\sqrt{\mu^2 - \omega^2}} = \bar{n} . 
\eeq
As we are interested in weak-coupling solution, we now let $\omega \rightarrow \epsilon \omega$ and $\mu \rightarrow \epsilon \mu$ (where $\epsilon$ is an order-counting parameter) to obtain the hydrogenic spectrum
\beq
\omega = \mu \left(1  - \frac{\alpha^2}{2 \bar{n}^2} \epsilon^2 + O(\epsilon^4) \right) . \label{eq:hydrogenic}
\eeq
Here $\alpha \equiv M \mu - Q q$ plays the role of an effective fine-structure constant, and the effective Bohr radius for the bound states is 
\beq
\label{ebr} a_0 = \frac{1}{\bar{n} k} = \frac{1}{\mu \alpha} . 
\eeq
The state with principal number $\bar{n}$ has an effective radius of $\bar{n}^2 a_0$. 

In principle, fine and hyperfine-structure terms can be calculated for the ABG bound state spectrum \cite{Baumann:2019eav, Cannizzaro:2023jle}; but this calculation is not pursued here. 

\subsubsection{The growth rate\label{subsec:growth}}
In general, quasibound states have complex frequencies, $\omega = \omega_R + i \omega_I$, due to the presence of the horizon. As can be seen from the time dependence in Eq.~(\ref{PHI}), $e^{-i \omega t} = e^{- i \omega_R t} e^{\omega_I t}$, the imaginary part $\omega_I$ determines whether the state is decaying ($\omega_I < 0$) or growing ($\omega_I > 0$) with time. In the special case $\omega_I = 0$, the quasibound state is stationary, and thus a true bound state. Such true bound states exist at the superradiant threshold with $\omega = \omega_c$, and they were recently studied in Ref.~\cite{Hod:2024aen}.

The imaginary part cannot be inferred from the far-field expansion alone, since the details of absorption and superradiance depend on the near-horizon region. In principle, one can carry out a matched asymptotic expansion, as in Ref.~\cite{Detweiler:1980uk} (N.B.~with a factor-of-two correction in Refs.~\cite{Pani:2012bp,Bao:2022hew}), to derive an expression for $M\omega_I$ in the $M \mu \ll 1$ regime, but this calculation is not pursued here. Nevertheless, based on that prior work \cite{Detweiler:1980uk,Pani:2012bp,Bao:2022hew}, we expect the imaginary part to grow as a power law, with the following scaling, 
\beq
M \omega_I \sim  \left( \omega - \omega_c \right) A_{\ell n} \alpha^{5 + 4 \ell}  , \label{eq:power-law}
\eeq
where $A_{\ell n}$ is a numerical factor, and $\alpha$ was defined in Eq.~(\ref{eq:alpha}). We test this conjecture in Sec.~\ref{sec:results}.

\subsection{Numerical method\label{subsec:numerical-method}}
\subsubsection{Expansion at the horizon}
Near the (outer) horizon, the ingoing solution can be expanded in the form of a Frobenius series. We start with the observation that the metric function $f(r)$ and the potential $\phi(r)$ can be expanded in series form,
\begin{subequations}
\begin{align}
f(r) &= \, 0 \, + \, \sum_{k = 1}^\infty \frac{1}{k!} f_k \, (r - r_+)^k ,  \\
\phi(r) &= \phi_+ + \sum_{k=1}^\infty \frac{1}{k!} \phi_k \, (r - r_+)^k, 
\end{align}
\label{eq:fphi-series}
\end{subequations}
where the expansion coefficients $f_k$ and $\phi_k$ are straightforward to obtain by differentiating Eq.~(\ref{eq:fABG}) and (\ref{eq:phiABG}) before evaluating numerically at $r=r_+$. The radial function $u(r)$ takes the form of a Frobenius series,
\begin{align}
u(r) &= (r - r_+)^\sigma \sum_{k=0}^\infty u_k \, (r - r_+)^k . \label{eq:u-series}
\end{align}
where $u_k$ are expansion coefficients to be determined. 
Inserting Eqs.~(\ref{eq:fphi-series}) and (\ref{eq:u-series}) into Eq.~(\ref{eq:radial}), and expanding around $r=r_+$, we obtain from the leading-order term 
\beq
\sigma = - \dfrac{i \tilde{\omega}}{f_1} .
\eeq
Without loss of generality, we set $u_0 = 1$, and then the sub-leading terms in the expansion of the radial equation determine the coefficients $u_{k \ge 1}$. At next order,
\begin{align}
u_1 &= \frac{\ell (\ell+1) + r_+ (f_1 + \mu^2 r_+)}{r_+^2 \alpha_1} + \frac{\tilde{\omega}}{2 f_1^2} \left( i f_2 + \frac{4q \phi_1 f_1}{\alpha_1}  \right)
\end{align}
where $\alpha_1 \equiv f_1 - 2 i \tilde{\omega}$. It is straightforward to obtain higher-order terms with a symbolic algebra package. 

A series expansion, truncated at order $N$, is used for initial data near the horizon, at $r = r_+ + \epsilon$. Typical values chosen are $N=4$ and $\epsilon = 10^{-3} M$. 

\subsubsection{Locating quasibound states}
A range of numerical methods have been applied to the challenge of computing the spectrum of quasibound states \cite{Cardoso:2005vk,Dolan:2007mj,Rosa:2011my,Frolov:2018ezx,Percival:2020skc}. Here we employ a simple but effective approach, namely, direct integration of the radial equation with numerical minimization. Starting with the series expansion (\ref{eq:u-series}), we integrate Eq.~(\ref{eq:radial}) from the near-horizon region to a sufficiently large radius (typically, $r_\text{max} \sim 800 / M \mu$). The quasibound state frequencies are located by seeking the local minima of $\log \left| u(r_{\text{max}}) \right|$ in the complex frequency domain. Starting values for $\omega$ are provided by the hydrogenic approximation, Eq.~(\ref{eq:hydrogenic}). 

\section{Results\label{sec:results}}
In this section we present a representative sample of quasibound states and their spectra from across the overlap region of the parameter space in Fig.~\ref{fig:parameter-space}. 

\newtext{\subsection{Radial profile and growth rate}}

\begin{figure}[!htbp]
    \includegraphics[width=1.0\columnwidth]{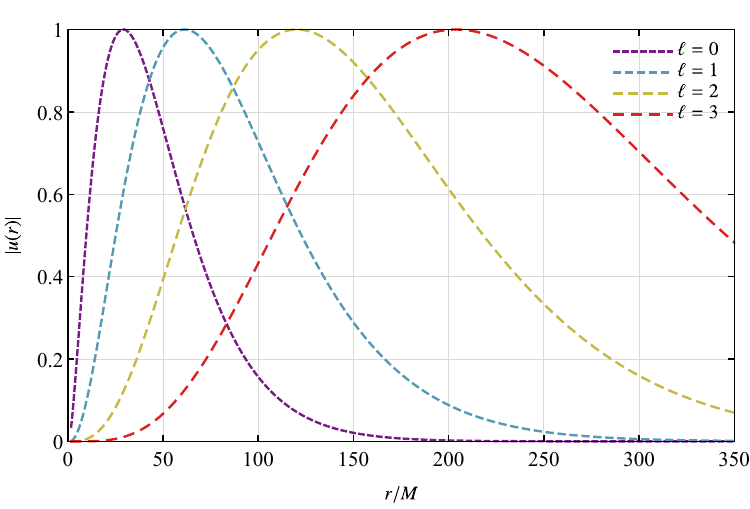}
    \includegraphics[width=1.0\columnwidth]{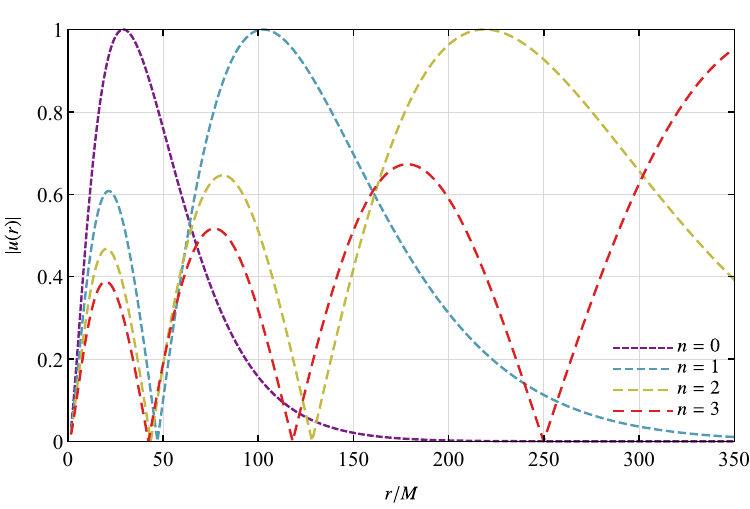}
    \caption{
     Example of bound state radial profiles for $Q/M = 0.4$, $M \mu / Qq = 1.5$, and $M \mu = 0.5$, as functions of $r/M$. Top: $\ell = 0, 1, 2, 3$ and $n=0$. Bottom: $\ell = 0$ and $n = 0, 1, 2$. 
    }
    \label{fig:wavefunctions_n0}
\end{figure}

Figure \ref{fig:wavefunctions_n0} shows the typical radial profile of bound states. The upper plot shows the fundamental mode for a range of $\ell$, and the lower plot shows the fundamental state ($n=0$) and two excited states ($n=1$, $n=2$) for $\ell = 0$. For small $M\mu$, the profiles are approximately hydrogenic (see Eq.~(\ref{eq:hydrogenic-wf})), with an effective radius of order $a_0 \bar{n}^2$, where $a_{0}$ 
is the effective Bohr radius of this system (see Eq.~\eqref{ebr}) and $\bar{n} = n + \ell + 1$ is the principle quantum number. 

\begin{figure}[!htbp]
    \includegraphics[width=1.0\columnwidth]{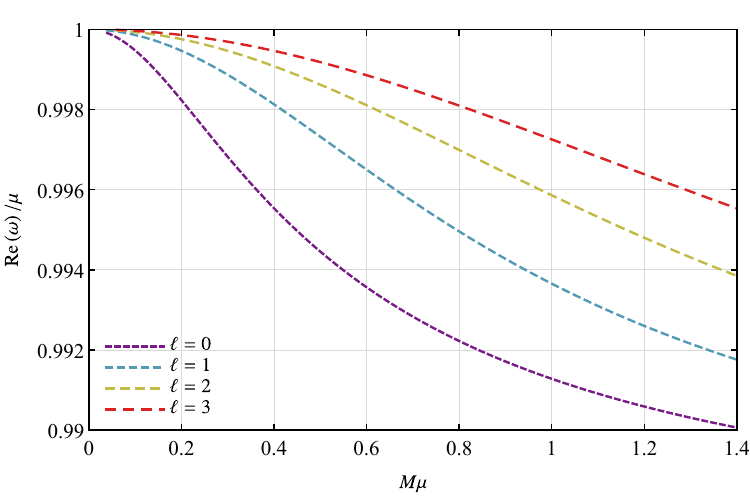}
    \includegraphics[width=1.0\columnwidth]{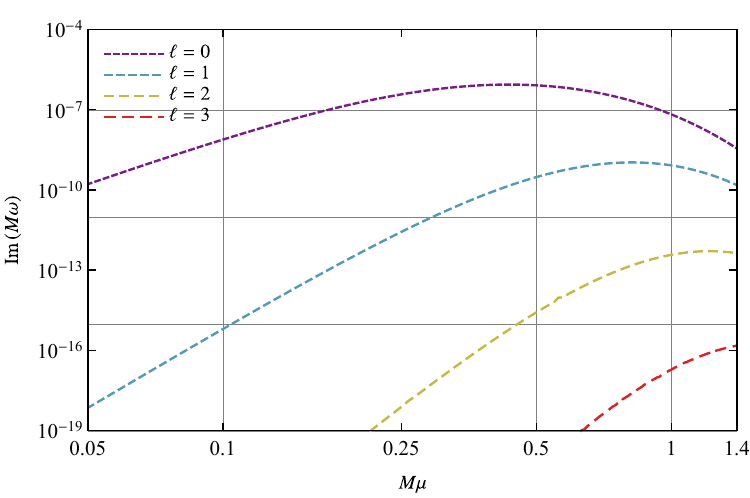}
    \caption{
     Spectrum of the fundamental superradiant bound state for $\ell = 0 \ldots 3$, as a function of $M \mu$. Here $Q/M = 0.4$, $M \mu / Qq = 1.5$. Top: The mode oscillation frequency ($\text{Re}(\omega)$). Bottom: The instability growth rate ($\text{Im}(\omega)$).
    }
    \label{fig:spectrum_l0123}
\end{figure}

Figure \ref{fig:spectrum_l0123} shows the spectrum of the fundamental ($n=0$) bound state for the lowest multipoles $\ell$, at fixed charge-to-mass ratios $Q/M = 0.4$ and $q / \mu = 0.6$, as a function of $M \mu$. At small $M \mu$, the real part is well approximated by the hydrogenic spectrum in Eq.~(\ref{eq:hydrogenic}). The ordering of energy levels with $\ell$ is maintained as $M \mu$ increases. All of the states shown are in the superradiant regime. As a function of $M \mu$, the growth rate $\text{Im} ( M \omega )$ increases in power-law fashion for small $M\mu$. The power-law index increases with $\ell$ in a manner consistent with Eq.~\eqref{eq:power-law}. 
The growth rate reaches a maximum at a certain value of $M \mu$, and then decreases from there. The same pattern is seen for each $\ell$. The $\ell = 0$ mode is the fastest growing, and thus it dominates the instability.

\begin{figure}[!htbp]
    \includegraphics[width=1.0\columnwidth]{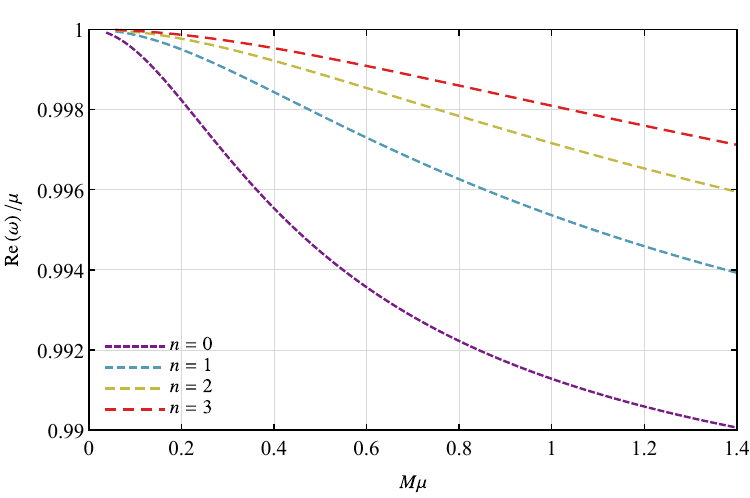}
    \includegraphics[width=1.0\columnwidth]{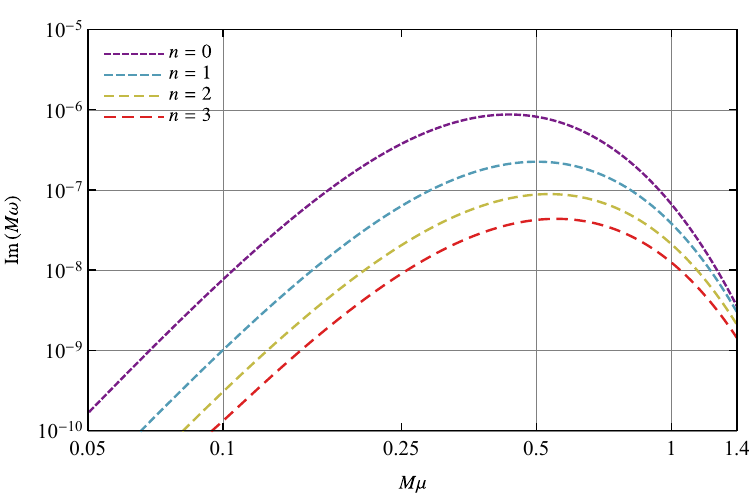}
    \caption{
     Spectrum of excited states ($n=0,1,2$) for $\ell = 0$, as a function of $M \mu$. Here $Q/M = 0.4$, $M \mu / Qq = 1.5$. Upper: The mode oscillation frequency. Lower: The instability growth rate.
    }
    \label{fig:spectrum_n012}
\end{figure}

Figure \ref{fig:spectrum_n012} shows the spectrum of the excited modes ($n = 0, 1, 2$) for $\ell = 0$. Again, the spectrum is consistent with the hydrogenic approximation at low $M\mu$, and the growth rate increases with $M\mu$ in a power law fashion. The modes $n = 0, 1, 2$ appear to have the same power-law index (but differing coefficients $A_{\ell n}$), which again is consistent with Eq.~(\ref{eq:power-law}). The fundamental mode ($n=0$) has the largest binding energy and the fastest growth rate, as expected.

\begin{figure}[!htbp]
    \includegraphics[width=1.0\columnwidth]{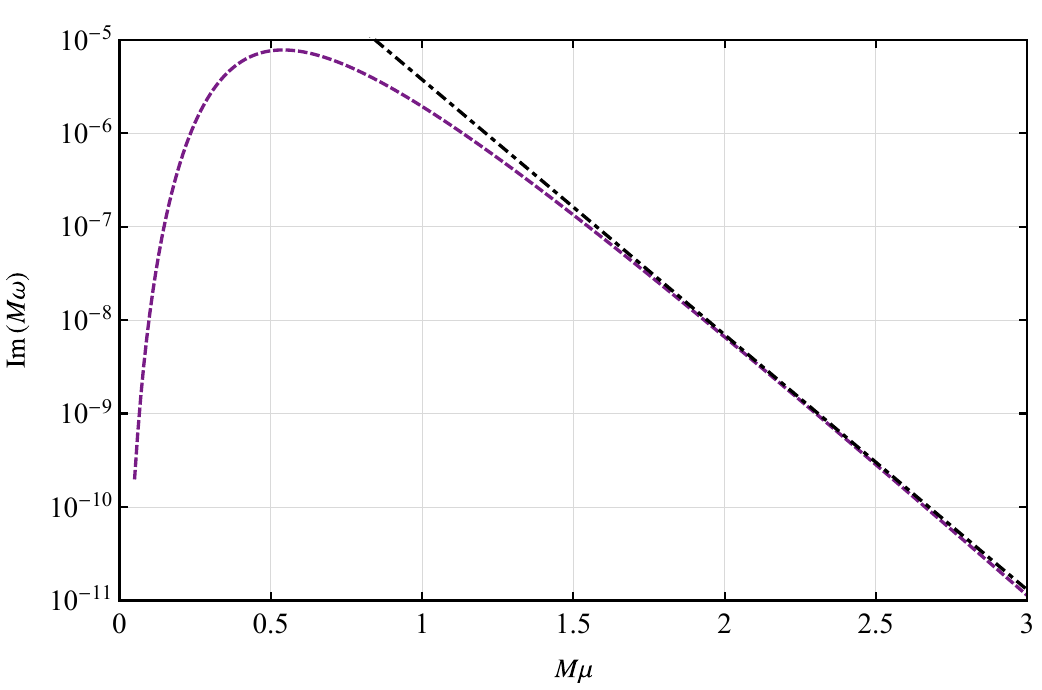}
    \caption{The growth rate of the monopole mode at large $M \mu$ on a semi-log scale, for parameters $Q/M = 0.63$, $M \mu / Q q = 2.2$ and $\ell = n = 0$, as a function of $M \mu$. The guideline (\newtext{dot-dashed line}) is $2 \times 10^{-3} \exp(-2 \pi M \mu)$.}
    \label{fig:l0_n0_im_exp}
\end{figure}

At large $M\mu$ and fixed charge-to-mass ratio, the numerical results indicate that the growth rate appears to fall off in exponential fashion with $M\mu$. This is illustrated in Fig.~\ref{fig:l0_n0_im_exp} for the particular case $Q/M = 0.63$, $M \mu / Q q = 2.2$ and $\ell = n = 0$. Exponential fall-off suggests that the near-horizon region is separated from the potential minimum by a potential barrier (in which case, the exponential index can be determined from a standard quantum-tunnelling argument, in principle \cite{Zouros:1979iw}).

\newtext{\subsection{Exploring the bound state spectrum of the ABG regular black hole}}

\begin{figure*}[!htbp]
    \includegraphics[width=1.0\columnwidth]{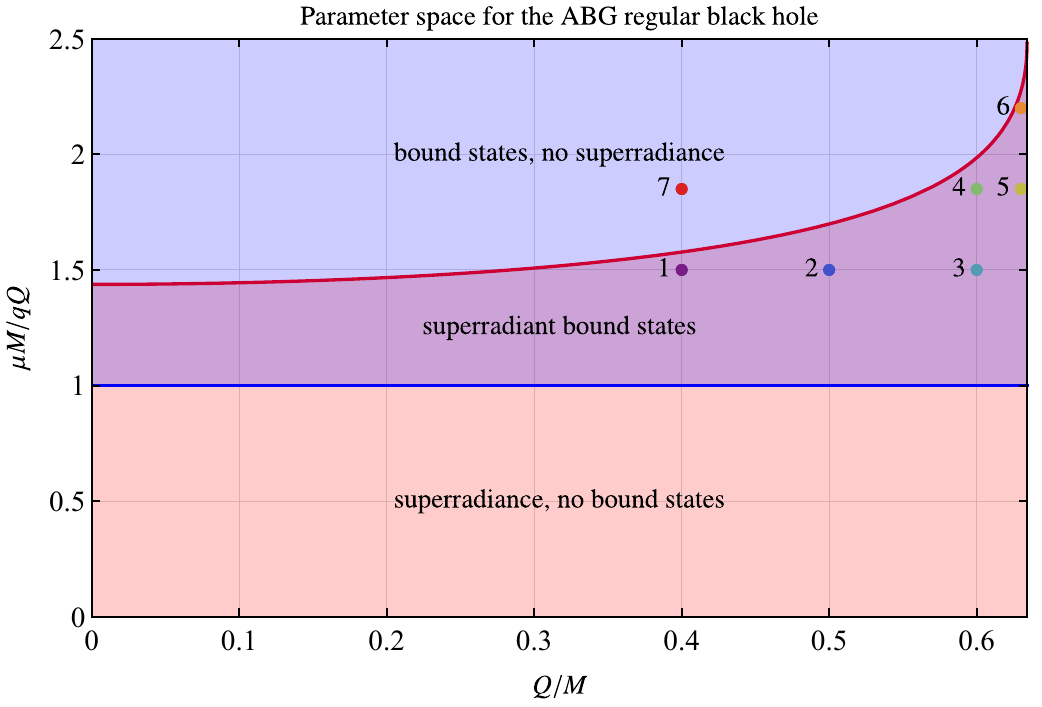}
    \includegraphics[width=1.0\columnwidth]{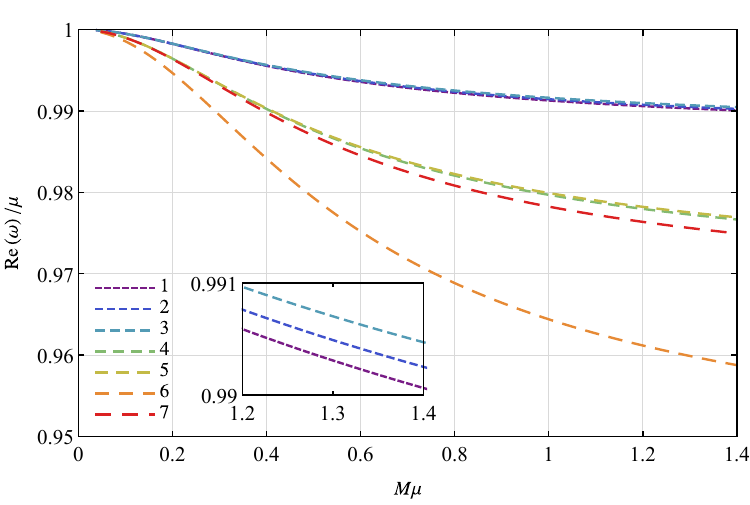}
    \includegraphics[width=1.0\columnwidth]{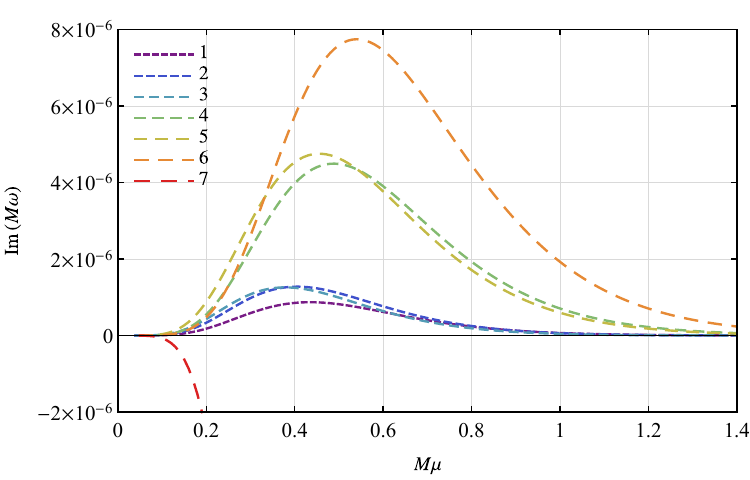}
    \includegraphics[width=1.0\columnwidth]{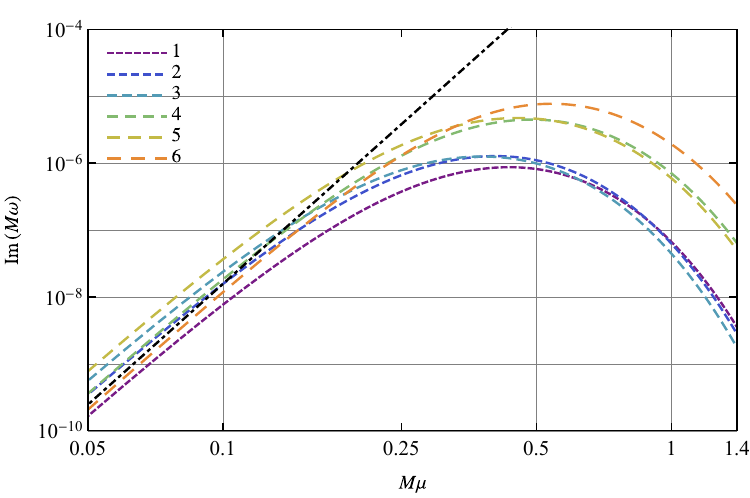}         
    \caption{
     Bound state spectrum for an ABG black hole, showing $\ell = 0$ and $n=0$ modes for fixed choices of $Q/M$ and $q/\mu$. Top left: the parameter space. The numbered points show the choices of charge-to-mass ratio for the black hole ($Q/M$) and the field ($q/m$). Top right: the real part of the bound state frequency as a function of $M\mu$. For small $M\mu$, the spectrum is approximately hydrogenic: see Eq.~(\ref{eq:hydrogenic}). \newtext{The inset shows the small differences in the frequency for the cases $1$--$3$.} Bottom left: the growth rate of the bound states for six cases inside the superradiant regime (labelled 1--6), and the decay rate for one case outside the superradiant regime (\newtext{red}, 7), as a function of $M \mu$. The fastest-growing mode shown here (\newtext{orange}, 6) corresponds to the parameter choice $Q/M = 0.63$ and $\mu M / q Q = 2.2$. Bottom right: the growth rate shown a log-log scale, as a function of $M \mu$. For small $M\mu$, the growth rate increases as a power of $M\mu$ (see Eq.~(\ref{eq:power-law})). The guideline (\newtext{dot-dashed black line}) is $(M\mu / 2)^6$. 
    }
    \label{fig:l0_n0}
\end{figure*}

Figure \ref{fig:l0_n0} shows the bound state spectrum of the fundamental monopole state ($\ell = n = 0$), as a function of $M\mu$, for several choices of $Q/M$ and $M \mu / qQ$ in the overlap region. As anticipated from the hydrogenic approximation (\ref{eq:hydrogenic}), the real part of the frequency depends principally on the effective fine-structure constant $\alpha$ (defined in Eq.~(\ref{eq:alpha})), and so modes on the same horizontal line in the parameter space (Fig.~\ref{fig:l0_n0}, top left) are grouped together (with groups $\{1,2,3\}$, $\{4,5,7\}$ and $\{6\}$ clear in Fig.~\ref{fig:l0_n0}). A similar grouping is clear for the growth rate. The power-law growth of $\text{Im}(M \omega)$ with $M\mu$ visible in Fig.~\ref{fig:l0_n0} (bottom right) is consistent with Eq.~(\ref{eq:power-law}). The mode outside the superradiant regime $\{7\}$ has a negative imaginary part (i.e.~it is decaying with time). The fastest-growing mode shown, $\{6\}$, is in the top-right region of the parameter space (i.e. it has the largest value of $M \mu / Q q$), as expected. 

\begin{figure*}[!htbp]
    \includegraphics[width=1.0\columnwidth]{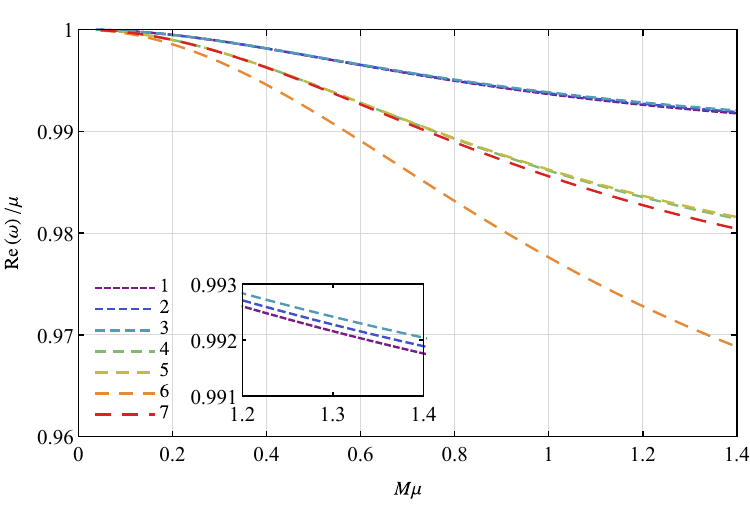}
    \includegraphics[width=1.0\columnwidth]{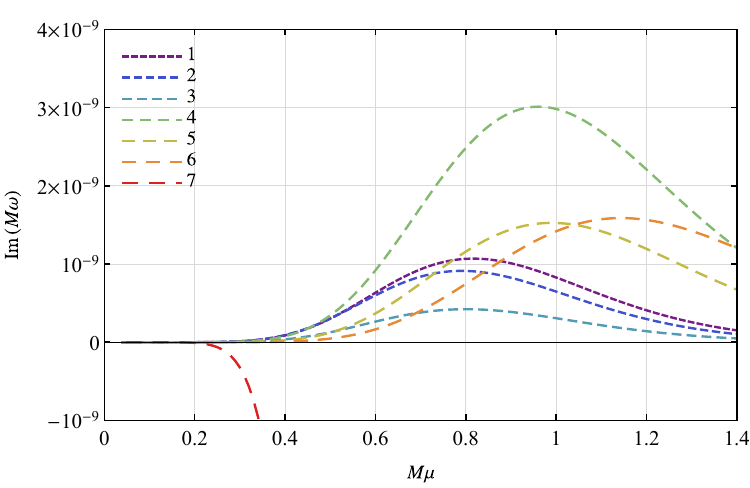}
    \includegraphics[width=1.0\columnwidth]{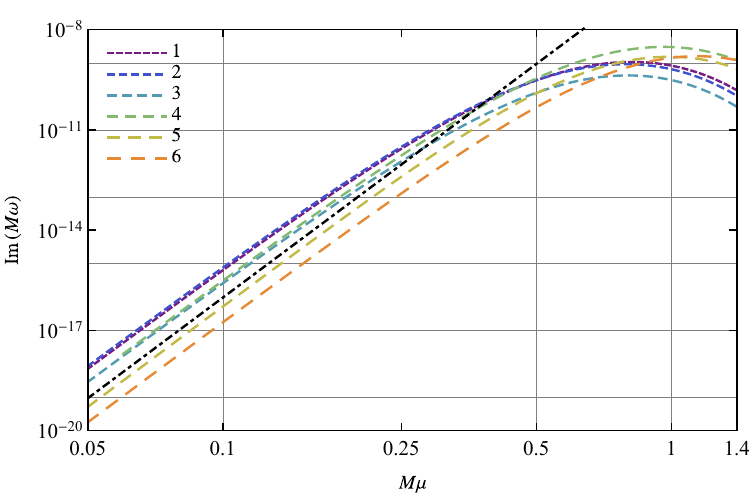}         
    \caption{Bound state spectrum of dipole modes, as a function of $M \mu$. As Fig.~\ref{fig:l0_n0}, but for the $\ell = 1$, $n=0$ bound states. The growth rate of the dipole ($\ell=1$) modes is significantly lower than the growth rate of the monopole ($\ell=0$) modes, by more than three orders of magnitude.  
    \newtext{For $\ell=1$, the mode with $Q/M = 0.6$ and $\mu M / qQ = 1.85$ (green, 5) has a larger maximum growth-rate than the mode with $Q/M = 0.63$ and $\mu M / qQ = 2.2$ (orange, 6). The guideline shows the power law $(M\mu / 4)^{10}$.} 
    }
    \label{fig:l1_n0}
\end{figure*}

Figure \ref{fig:l1_n0} shows the bound state spectrum of the fundamental dipole state ($\ell = 1$, $ n = 0$), for the same parameter choices as in Fig.~\ref{fig:l0_n0}. Comparing the $\ell = 0$ and $\ell=1$ results, we observe in the latter a steeper power-law for the growth rate (consistent with Eq.~(\ref{eq:power-law}) with an index $10$ instead of $6$); a substantially smaller maximum growth rate, also seen in Fig.~\ref{fig:spectrum_l0123} ($3 \times 10^{-9}$ vs $8 \times 10^{-6}$); and ordering of the real part of $\omega$ in accordance with the hydrogenic approximation, Eq.~(\ref{eq:hydrogenic}), in both cases. Intriguingly, for $\ell = 1$ the fastest growth rate arises for the parameter choices marked (4) in Fig.~\ref{fig:l0_n0}, whereas for $\ell = 0$ the fastest growth rate is for the choice marked (6). The former corresponds to a lower black hole charge-to-mass ratio ($Q/M=0.6$) than the latter ($Q/M=0.63$).

\begin{figure}[!htbp]
    \includegraphics[width=1.0\columnwidth]{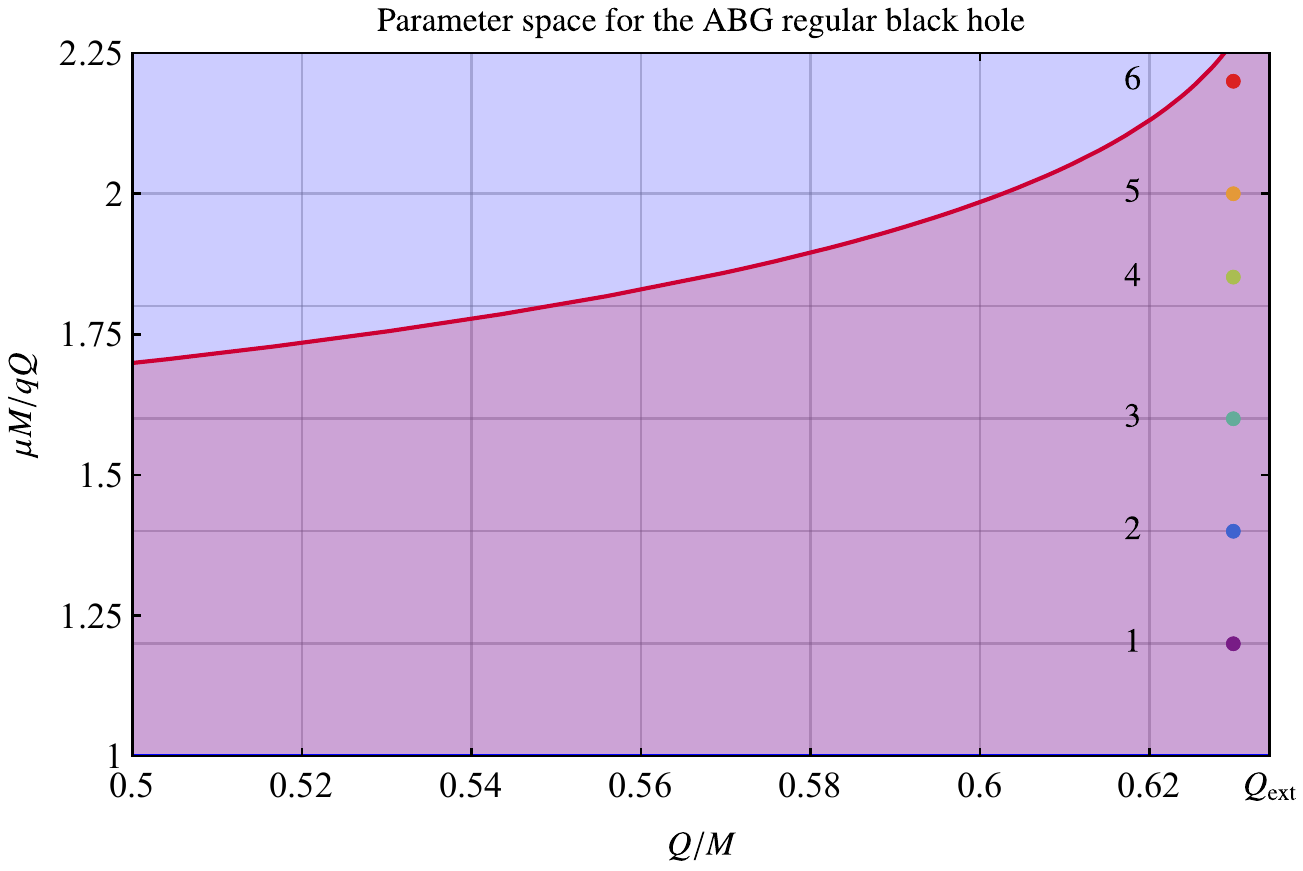}
    \includegraphics[width=1.0\columnwidth]{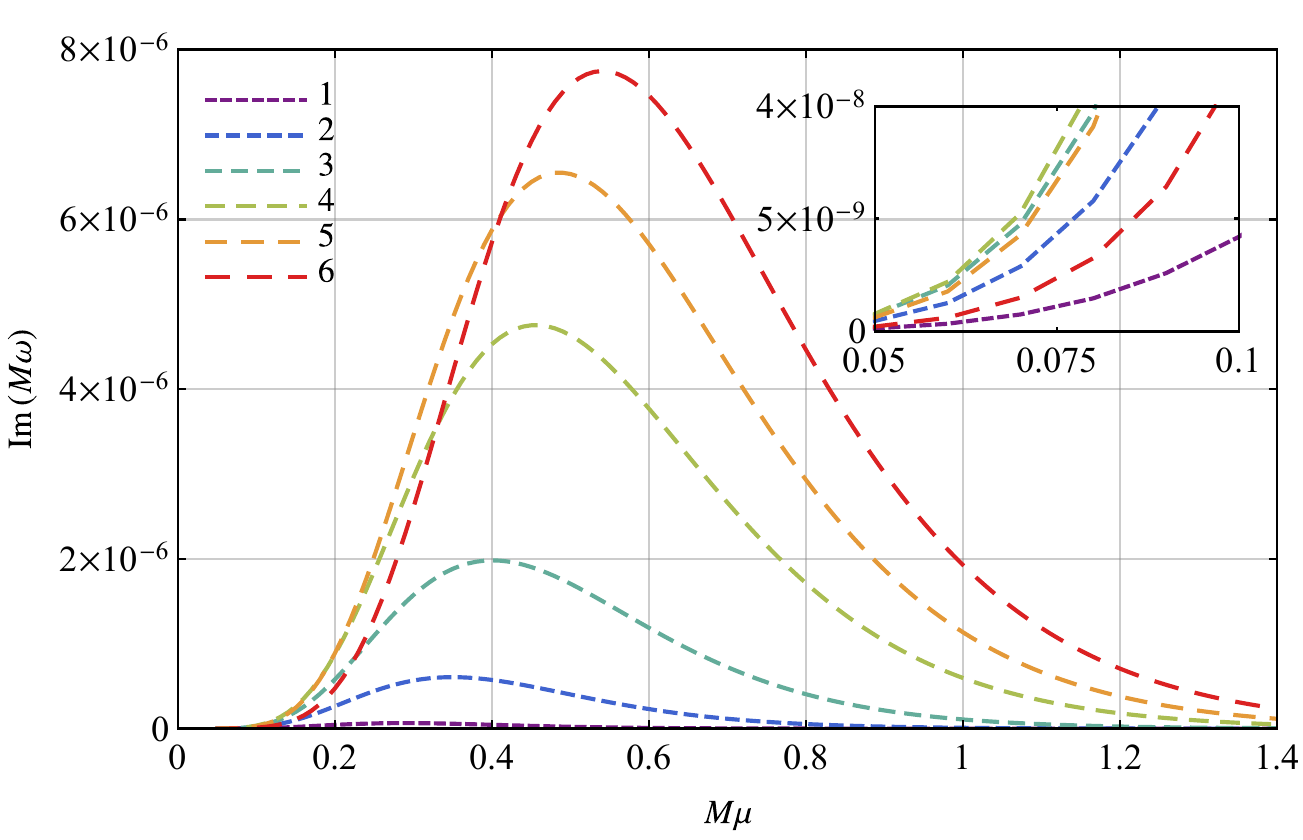}
    \caption{\newtext{The superradiant instability considering different values of $\mu M/qQ$ for fixed choices of $Q/M$. Upper: The parameter space for $Q/M = [0.5, Q_{\rm{ext}}]$. Lower: The instability growth rate of certain configurations labeled in the parameter space, as functions of $M \mu$. The inset helps to visualize the instability growth for $\mu M \ll 1$.}}
    \label{fig:diffalpha}
\end{figure}

\newtext{It is also interesting to examine the asymptotic behavior of the instability growth for large values of $Q/M$. Figure~\ref{fig:diffalpha} displays the instability growth rate at certain points on the parameter space. For large values of $\mu M$, we note that higher values of $\mu M/qQ$ (but below the superradiant cutoff) create a faster instability. On the other hand, for small values of $\mu M$ at fixed $Q/M$, we observe that the behavior of the instability is more subtle. The parameter choice $\mu M/qQ$ with the highest maximum growth rate will not necessarily have the highest growth rate at low $M\mu$ [see, e.g., points 2 (blue dashed line) and 6 (purple dashed line) and the inset]. Furthermore, in the limit $\mu M \ll 1$, the rate does not appear to grow (or fall) linearly with the chosen values of $\mu M / qQ$ at fixed $Q/M$ (notice that the curves in the bottom panel satisfy the following order: 1, 6, 2, 5, 3, 4).
A possible explanation for this ordering is contained in Eq.~(\ref{eq:power-law}): at fixed $Q/M$, parameter choices $\mu M / q Q$ closer to the superradiant cutoff (e.g.~point 6 in Fig.~\ref{fig:diffalpha}) correspond to larger values of $\alpha$, but smaller values of $\omega - \omega_c$.
}

\begin{figure}[!htbp]
    \includegraphics[width=1.0\columnwidth]{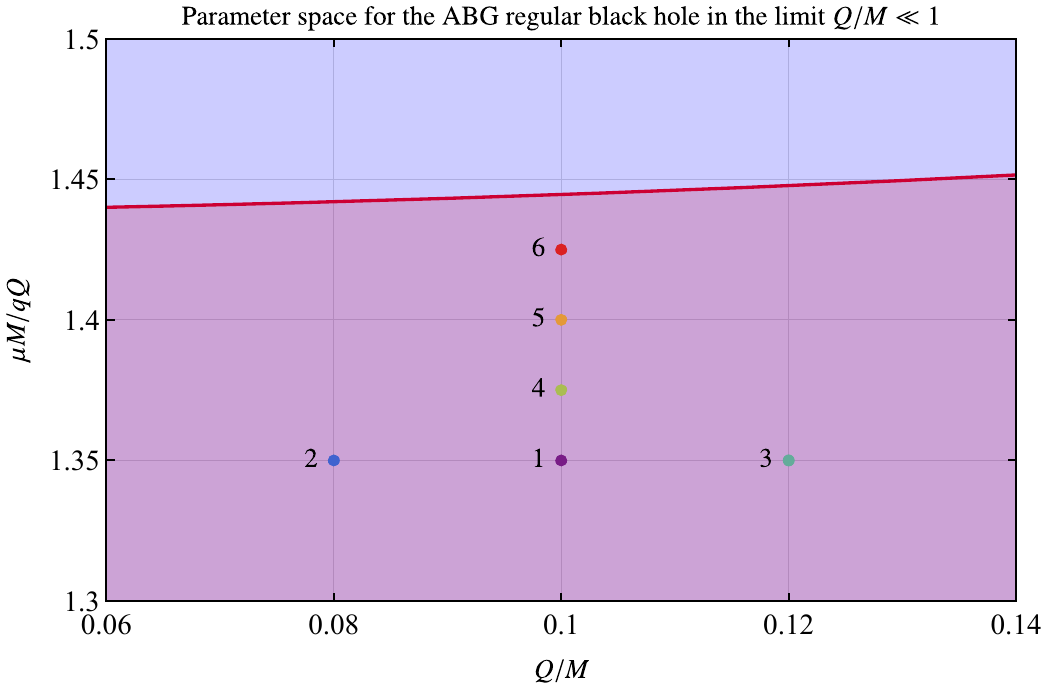}
    \includegraphics[width=1.0\columnwidth]{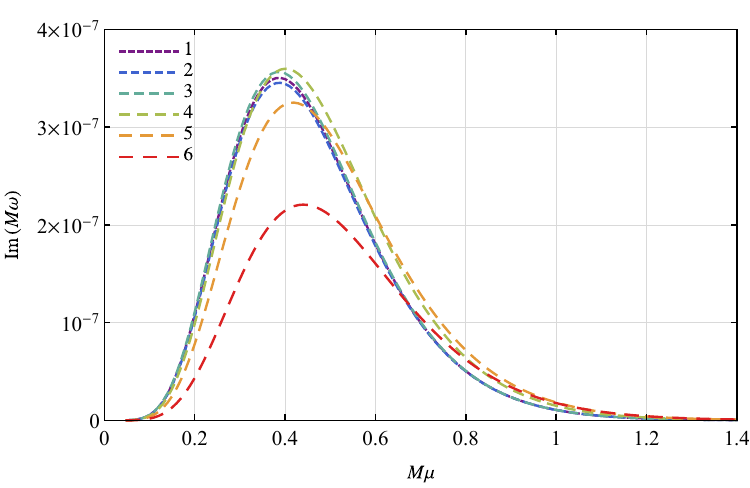}
    \caption{
     The superradiant instability at low charge-to-mass ratios $Q/M$. Upper: the parameter space for $Q/M = [0.06, 0.14]$. Lower: The instability growth rate of certain configurations labeled in the parameter space, as functions of $M \mu$.
    }
    \label{fig:lowQ}
\end{figure}

An interesting facet of the superradiant instability on the ABG spacetime is that it remains significant even for black holes with a small charge-to-mass ratio ($Q/M \ll 1$). Figure \ref{fig:lowQ} shows the growth rate for $Q/M \approx 0.1$, demonstrating that it can still work quite efficiently ($\text{Re}(M \omega) \sim 3.5 \times 10^{-7}$ at $M \mu \approx 0.4$) if the charge-to-mass ratio of the \emph{field} is compensatingly large ($q / \mu \approx 7.3$ in this case). 

\begin{figure}[!htbp]
    \includegraphics[width=1.0\columnwidth]{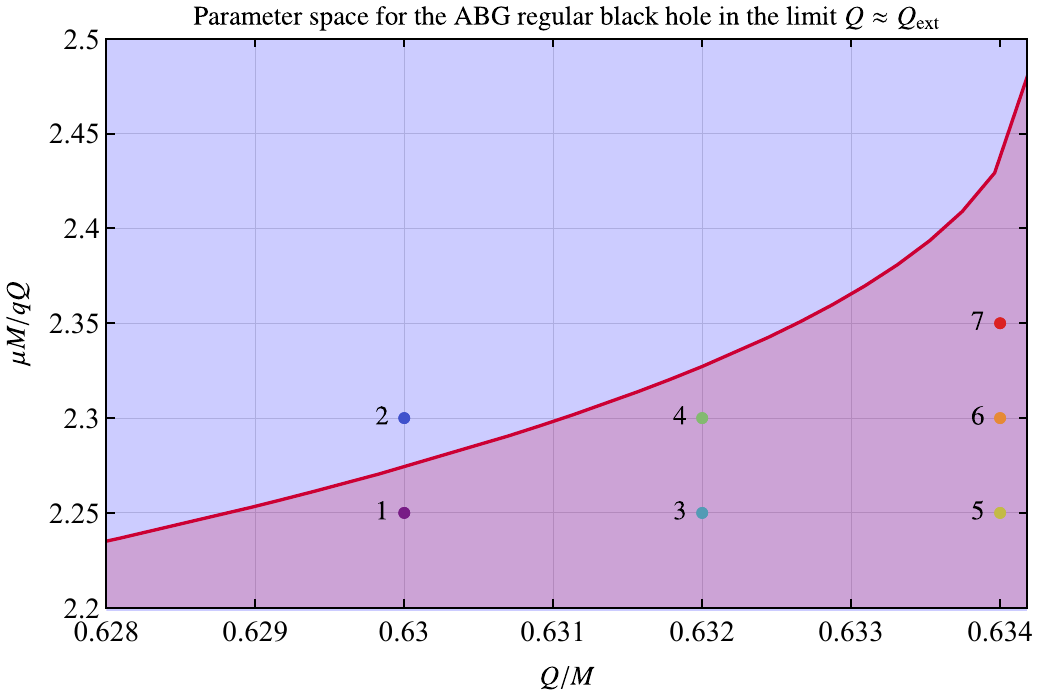}
    \includegraphics[width=1.0\columnwidth]{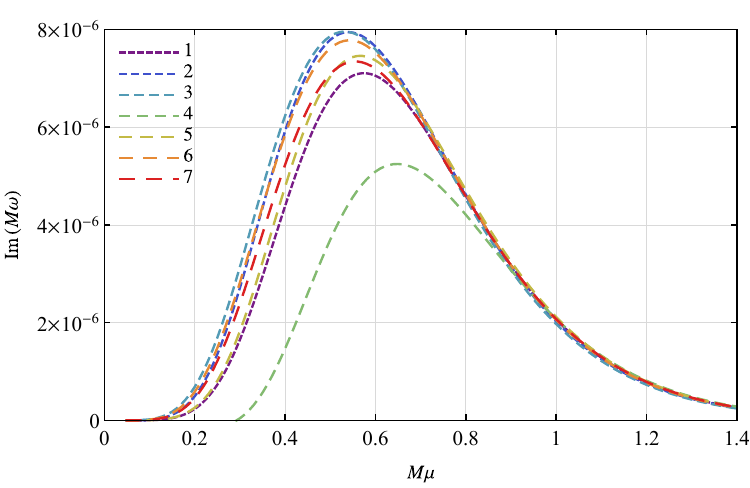}
    \caption{
     The superradiant instability at high charge-to-mass ratios $Q/M$. Upper: the parameter space for $Q/M = [0.628, 0.634]$. Lower: The instability growth rate for the corresponding configurations highlighted in the upper plot, as functions of $M \mu$.
    }
    \label{fig:highQ}
\end{figure}

The fastest-growing instabilities occur for black holes carrying significant charge, with charge-to-mass ratios near the extremal bound, $Q \lesssim Q_c \approx 0.63418$. Figure \ref{fig:highQ} shows growth rates in this regime. The results are consistent with a maximum growth rate of $\text{Im} (M \omega) \approx 8 \times 10^{-6}$ for the monopole mode. For comparison, in the Kerr black hole, considering bosonic fields, we have $\text{Im} (M \omega) \approx 1.72 \times 10^{-7}$~\cite{Dolan:2012yt}, and considering neutral massless scalar fields with a mirrorlike boundary condition, we have $\text{Im} (M \omega) \approx 6 \times 10^{-5}$~\cite{Cardoso:2004nk}. For RN black holes in a cavity, it is possible to obtain states with $\text{Im} (M \omega) \approx 0.07$~\cite{Herdeiro:2013pia}. Therefore, our results show that the unstable modes in the regular ABG black hole can grow faster than in the standard Kerr case but slower than in the scenarios with a reflecting mechanism (for scalar fields). We draw some more comparisons in detail in the conclusions.

Finally, we consider the stationary bound states with real frequencies at the superradiance threshold: $\omega = \omega_c$. In parameter space, the stationary bound states lie just above the superradiance-boundary curve in Fig.~\ref{fig:normal_mode} (since $\omega < \mu$), and they approach this curve as $M \mu \rightarrow 0$. These states were recently investigated in Ref.~{\cite{Hod:2024aen} where, through analytic arguments, it was shown that in the $Q / M \ll 1$ regime, the stationary bound states lie in the narrow range 
\beq
\dfrac{23}{16} \le \dfrac{M \mu}{Q q} \le \sqrt{\dfrac{6859}{3240}},
\eeq
with the upper (lower) bound corresponding to the limit $M\mu \rightarrow \infty$ ($M\mu \rightarrow 0$). This range is indicated in Fig.~\ref{fig:normal_mode} (grey line), and it appears to be consistent with our numerical results at larger values of $Q/M$.

\begin{figure}
    \includegraphics[width=1.0\columnwidth]{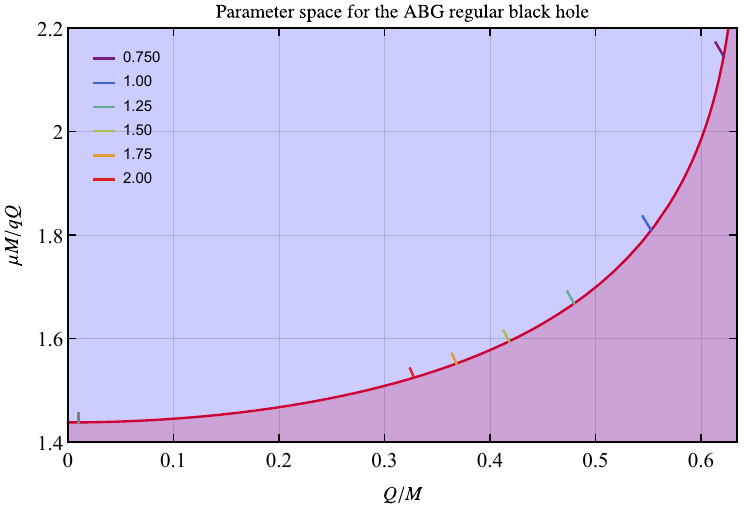}
    \caption{
     Stationary bound state spectrum. The short colored lines show the spectrum of states with real frequencies $\omega = \omega_c$, for field charge-to-mass ratios $q/\mu$ ranging from $0.75$ to $2$ (right-to-left), and $M\mu \in [0.05,2.0]$. These lines approach the approximate superradiant boundary (dashed purple line) from above as $M\mu \rightarrow 0$. The short grey line, left, shows the range $[23/16, \sqrt{6859/3240}]$ derived in Ref.~\cite{Hod:2024aen}, which is valid in the limit $Q/M \ll 1$ (i.e.~$q/\mu \gg 1$). 
    }
    \label{fig:normal_mode}
\end{figure}

\newtext{\subsection{Comparisons with the growth rate in the Kerr case}}
\newtext{Figure~\ref{fig:abgcompkerr} compares the instability growth rate of the massive and charged Klein-Gordon (scalar) field on the ABG background with the massive Klein-Gordon and Proca fields on the Kerr background (studied in Refs.~\cite{Dolan:2007mj} and \cite{Dolan:2018dqv}, respectively). For the ABG scenario, we take the same parameters as used in Fig.~\ref{fig:l1_n0}. For the Kerr case, we consider a rapidly-rotating black hole with $a/M = 0.99$ and the corotating mode $m = 1$. For the Proca field we have shown the ``odd-parity'' and ``even-parity" polarizations  $S = 0$ and $S = -1$, respectively ($S=+1$ is omitted for clarity).}

\begin{figure*}
		\includegraphics[width=1.0\columnwidth]{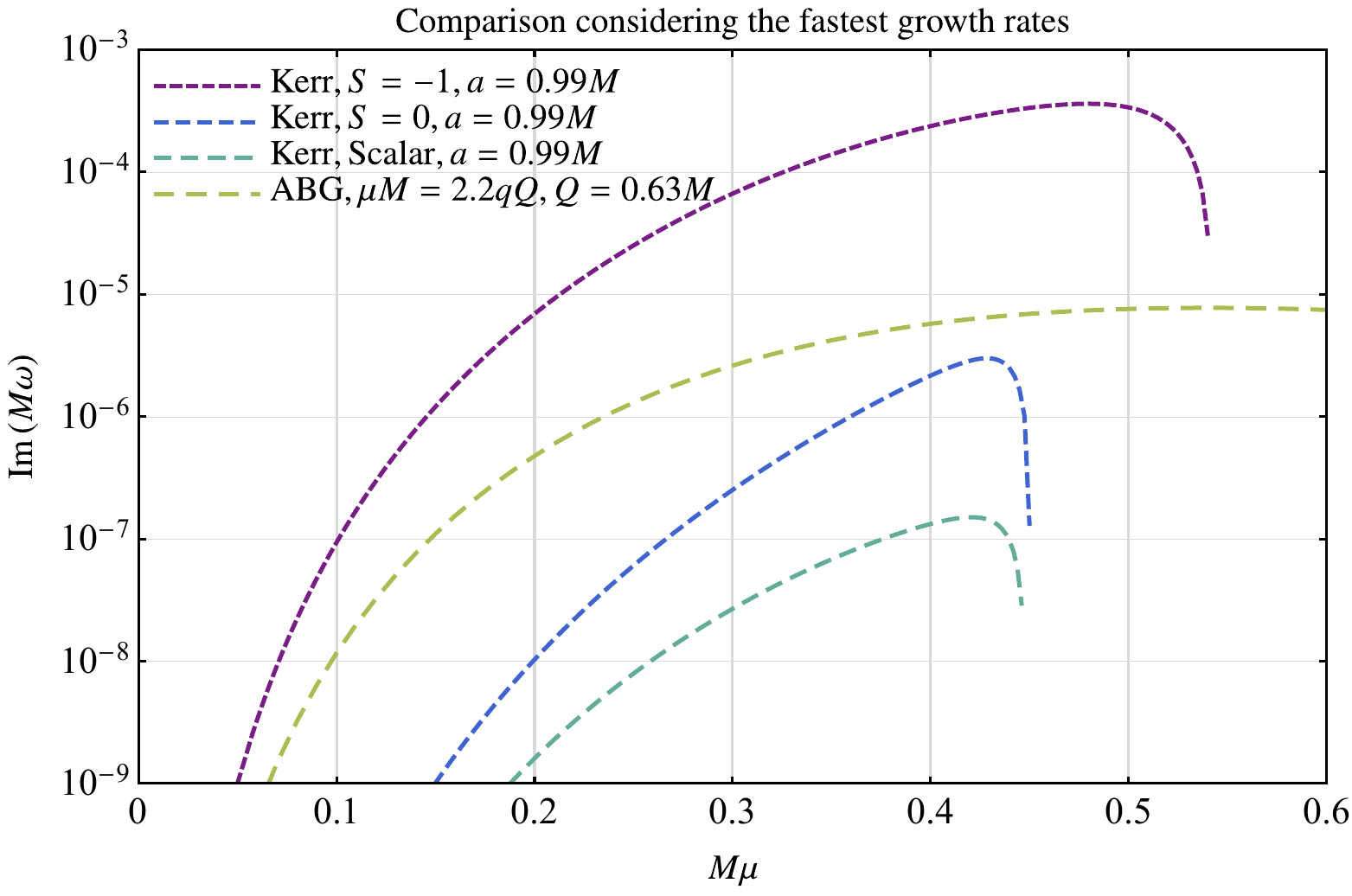}
	\caption{\newtext{Comparison between the growth rates of a charged and massive scalar field on an ABG background, and massive scalar and Proca fields on Kerr background.}}
	\label{fig:abgcompkerr}
\end{figure*}

\newtext{Comparing the scalar-field results, the instability is grows faster around a highly-charged regular BH than a rapidly-spinning Kerr BH. However, the Proca-Kerr scenario, the situation is more subtle. For the polarization $S = 0$ (and in the limit of higher values of $\mu M/qQ$ and $a/M$, for $ M \mu \approx 0.45$) the growth rate of the Proca field on a Kerr black hole is comparable in magnitude, but still somewhat slower than, the growth rate of a scalar field on the ABG spacetime. However, for the polarization $S = -1$ the Proca bound state has a radial profile more similar to that of the $\ell = 0$ scalar mode, and Proca field (around a Kerr BH) grows more rapidly than the scalar field (around an ABG BH), even in the limit of higher values of $\mu M/qQ$.}

It is also interesting to compare the \emph{maximum} possible growth rates in the parameter space. For a charged scalar field on the ABG spacetime, we here find a maximum growth rate (in the monopole $\ell=0$ sector) of $M \omega_I \approx 8 \times 10^{-6}$. This is significantly faster growth than is found in a scalar field on Kerr spacetime, for which the fastest growth rate is $M \omega_I \approx 1.72 \times 10^{-7}$ in the corotating dipole ($\ell = m = 1$) mode at $M\mu \approx 0.45$ and $a/M \approx 0.997$ \cite{Dolan:2012yt}. However, a Proca field on Kerr spacetime can grow at substantially faster rates \cite{Pani:2012bp,Baryakhtar:2017ngi,Cardoso:2018tly, Frolov:2018ezx, East:2017ovw}: $M \omega_I \approx 4.27 \times 10^{-4}$ at $M \mu \approx 0.542$ and $a/M \approx 0.999$ \cite{Dolan:2018dqv}. Moreover, a massive spin-two field demonstrates additional features: an unstable spherical mode, and a special dipole polar mode which does not have a hydrogenic spectrum, and which exhibits a rapid growth rate as large as $\text{Im} (M \omega) \approx 0.19$ at $M \mu = 0.8$ \cite{Dias:2023ynv}.

\vspace{0.2cm}


\section{Discussion and Conclusions\label{sec:conclusions}}

In the preceding sections, we have demonstrated that a charged, massive scalar field in the vicinity of an ABG black hole can undergo a superradiant instability. The results presented here are consistent with the partition of parameter space in Fig.~\ref{fig:parameter-space}, which highlights that the instability arises under the reasonably generic conditions that  both superradiance and quasibound states coexist. For fixed charge-to-mass ratios, the growth rate as a function of $M\mu$ is governed by a power law at low $M\mu$ (see Eq.~(\ref{eq:power-law}) and Figs.~\ref{fig:spectrum_l0123}, \ref{fig:spectrum_n012}, \ref{fig:l0_n0} and \ref{fig:l1_n0}) before reaching a maximum value and then decaying exponentially at high $M\mu$ (see Fig.~\ref{fig:l0_n0_im_exp}). The maximum growth rate can reach $\text{Im}(M \omega) \approx 8 \times 10^{-6}$ for the monopole state ($\ell = 0$) at $M \mu \approx 0.53$ (Fig.~\ref{fig:highQ}); the higher multipoles (Fig.~\ref{fig:spectrum_l0123}) and the excited states (Fig.~\ref{fig:spectrum_n012}) have slower growth rates. True (stationary) bound states exist at the threshold $\omega = \omega_c$, and these lie in a narrow band of parameter space (Fig.~\ref{fig:normal_mode}).

This work adds to a body of literature showing that superradiant instabilities (and hairy black holes) can arise in a variety of scenarios. To name but a few: in massive dynamical Chern-Simons gravity \cite{Richards:2023xsr}, in theories with self-interacting scalar fields \cite{Baryakhtar:2020gao, Chia:2022udn}, in modified-gravity theories with a scalar field coupled to curvature invariants \cite{HegadeKR:2022xij,R:2022tqa}, in standard GR with a coupled electromagnetic and axion field \cite{Burrage:2023zvk}, and scalar fields interacting with exotic compact objects \cite{Zhou:2023sps}. In this work, we have demonstrated that a charged instability afflicts an example of a \emph{regular} black hole in nonlinear electrodynamics. Several other solutions in this class have been proposed; it would be interesting to investigate which are (in principle) affected by superradiant instabilities. 

Several avenues for further work are open. These avenues include, first, a derivation of approximate expressions for the growth rate in the low-$M\mu$ and high-$M\mu$ regimes, extending the approaches of Detweiler \cite{Detweiler:1980uk}, and Zouros and Eardley \cite{Zouros:1979iw}, respectively. Second, an investigation of fields of higher spin on the ABG black hole. Third, a survey of regular black holes in nonlinear electrodynamics to classify which are afflicted by the instability in principle. Fourth, analytical study of the stationary bound states outside the $Q/M \ll 1$ regime, building on the work of Ref.~\cite{Hod:2024aen}, and their interpretation as the starting point of a non-linear hairy configurations \cite{Hod:2012px, Herdeiro:2014goa,Herdeiro:2015waa, Benone:2015jda}. Fifth, pursuit of the instability into the non-linear regime (where e.g.~the stress-energy of the field generates additional spacetime curvature), to assess whether the system approaches a stable hairy configuration (i.e.~a black hole embedded in a `cloud'), or whether it undergoes a cycle of explosive outflows (the so-called Bosenova \cite{Yoshino:2015nsa}). If the former is true, one would wish to assess how much of the black hole's mass can be deposited into the field. Sixth (and relatedly), an investigation of plausible scalar-hairy configurations linked to the ABG black hole, and their stability. 

\newtext{
In fact, it is relatively straightforward to make an estimate of an upper bound on the mass deposited into the field by the black hole. The instability drives the black hole parameters  along a dashed line shown in Fig.~\ref{fig:parameter-trajectories} (upper plot). It is quick to establish that the charge-to-mass ratio of the black hole is decreasing, and so the system progresses from right to left, from $\tilde{Q}_i \equiv Q_i / M_i$ to $\tilde{Q}_f \equiv Q_f / M_f$. We assume that the charge-to-mass ratio of the field itself is fixed, and this ratio determines the relative loss rates: $dQ/dM \approx q / \mu = \text{const}$. Now let us suppose that the instability starts at the point where bound states first form, when $q \tilde{Q}_i / \mu = 1$; and that it proceeds until the system reaches the superradiant cut-off at $\tilde{Q}_f$. It follows that
\newtext{\beq
\tilde{Q}_f = \dfrac{\tilde{Q}_i + (1/\tilde{Q}_i) (\Delta M / M_i)}{\left(1 + \Delta M / M_i \right)},
\eeq}
where $\Delta M = M_f - M_i$ is the change in the black hole mass. Rearranging, we come to a formula for (an approximate upper bound on) the percentage mass loss of
\beq
\frac{\Delta M}{M_i} = - \left( \frac{\tilde{Q}_i - \tilde{Q}_f}{\tilde{Q}_i^{-1} - \tilde{Q}_f} \right) .
\eeq
The lower plot in Fig.~\ref{fig:parameter-trajectories} shows the relative mass loss for the system as a function of its initial charge to mass ratio. For $\tilde{Q}_i = Q_c / M \approx 0.6341$ (the extremal case), we find $\tilde{Q}_f \approx 0.40167$ and an upper bound on the mass loss of
\newtext{\beq
\dfrac{|\Delta M|}{M_i} \approx 19.785\%.
\eeq}As a point of comparison, numerical studies have shown that a rotating black hole can transfer circa $9\%$ of its mass (and circa $38\%$ of its angular momentum) into a neutral Proca field \cite{East:2017mrj,East:2017ovw,East:2018glu}.
}

\begin{figure}
    \includegraphics[width=1.0\columnwidth]{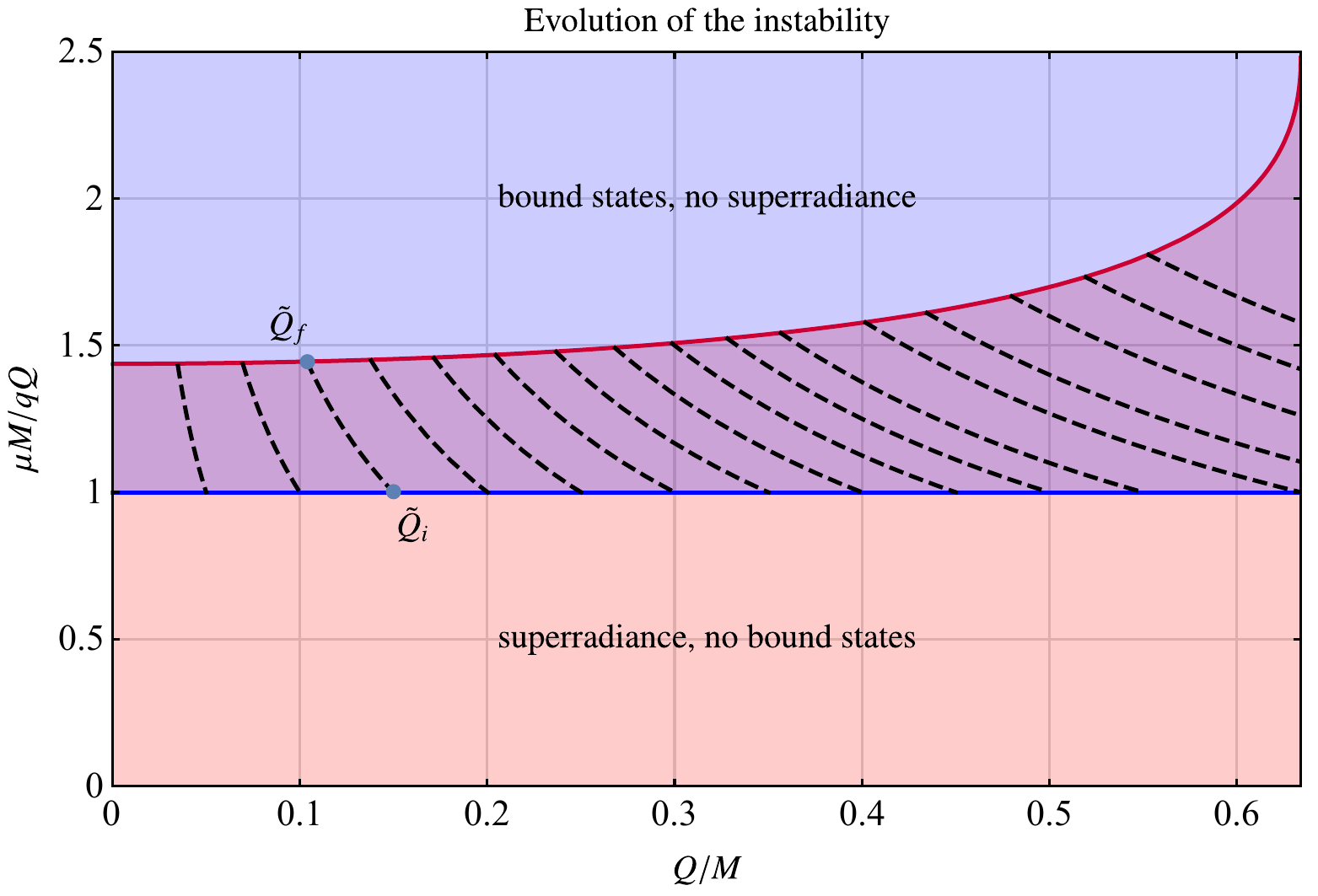} \\
    \vspace{0.3cm}
    \includegraphics[width=1.0\columnwidth]{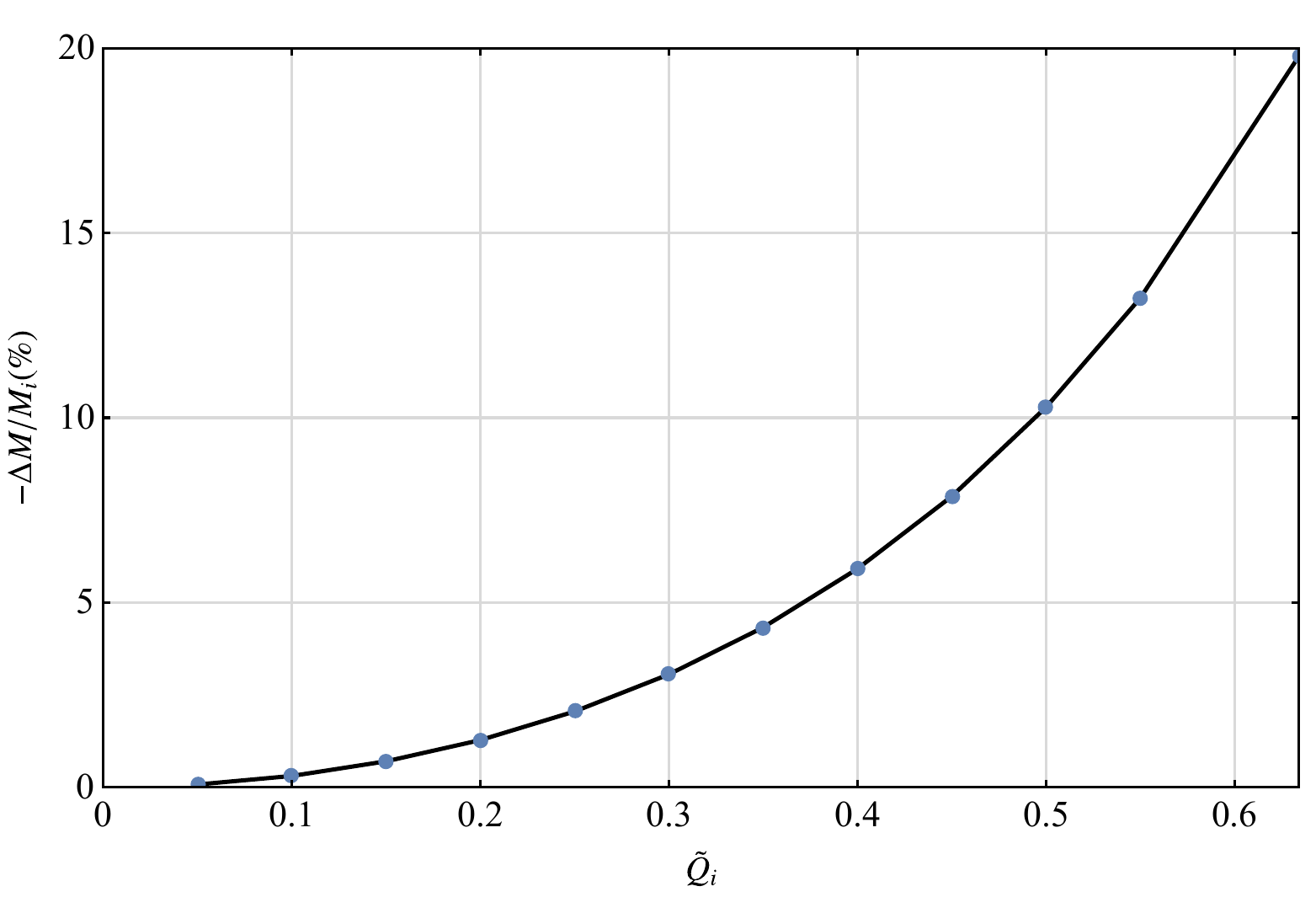}
    \caption{
     Upper plot: Under the instability, systems move from right to left along the dashed lines in parameter space. Lower plot: An estimate of (an upper bound on) the percentage of mass lost by the black hole over the full course of the instability, shown as a function of initial charge-to-mass ratio, $\tilde{Q}_i \equiv Q_i / M_i$.
    }
    \label{fig:parameter-trajectories}
\end{figure}

In the case of \emph{rotational} superradiance and the Kerr black hole, substantial work has gone into addressing the question of the non-linear development of the instability, and here a lack of symmetry necessitates expensive numerical simulations \cite{Witek:2012tr, Yoshino:2015nsa, East:2017mrj,East:2017ovw,Santos:2020sut}. By contrast, tracking a \emph{spherically-symmetric} perturbation is comparatively straightforward. This observation has led some authors to consider black holes and solitons in Einstein-Maxwell-scalar theory  but, as noted here, in \emph{linear} electromagnetism the RN black hole does not appear to admit superradiant bound states. Typically, this leads authors to the (perhaps artificial) device of adding a mirror or other boundary to confine the field \cite{Dolan:2015dha,Dias:2018zjg,Davey:2021oye}. Here we have shown that, in certain theories of nonlinear electrodynamics, this device is redundant, since a charged field with a mass can undergo a superradiant instability without imposing a mirror. 

Regular black holes are particularly intriguing subjects for study, because they appear to evade the breakdown of the theory at a spacetime singularity. However, the stability of such solutions within the outer horizon is an open question. A complete non-linear study of the superradiant instability will necessitate some consideration of dynamics in the interior of the black hole, including near the inner horizon. In this context, it is well-known that standard charged black holes in general relativity typically present the so-called mass inflation instability, characterised by exponential growth of the black hole mass function at the inner horizon. On the other hand, for certain regular black holes (e.g.~Hayward),  exponential growth is replaced by a polynomial growth, and the singularity is of a weaker integrable type \cite{Bonanno:2020fgp,Bonanno:2022jjp}. Moreover, in configurations in which the surface gravity of the inner horizon is zero, mass inflation is eliminated \cite{Carballo-Rubio:2022kad}. Hence, in principle, these geometries could provide a good scenario for a complete non-linear study of instabilities.

\newtext{We also highlight some potential observational consequences of our results. It is well known that ultralight bosons can trigger the superradiant instability in rotating black holes, spinning them down and leaving an exclusion region in the black hole mass-spin plane~\cite{Brito:2014wla,Arvanitaki:2016qwi,Zhou:2023sps}, and so, using measurements of the spin and mass of astrophysical black holes, one can constrain the mass of the scalar field. Assessing whether this is physically possible/reasonable in the context of regular charged black holes, considering charged scalar fields, is a potentially relevant problem. In other words, would it be possible to construct an exclusion region in the charge-mass plane of the regular black hole based on the superradiant instability of charged scalar fields? And how would this, as well as the superradiant instability per se, differ from the case of standard (i.e., singular) black holes or other compact objects? These questions are worthy of further study, as they allow us to explore some potential observational consequences of the superradiant instability in the context of regular black holes.}

We conclude with the conjecture that, under the superradiant instability, the exterior of the black hole will evolve towards a quasistable configuration with long-range spherically-symmetric scalar hair. Moreover, we conjecture that this non-linear configuration emerges from a smooth extension of the stationary bound states investigated here and in Ref.~\cite{Hod:2024aen}. In investigating stability, an important question to address will be, if both `hairy' and electrovacuum black holes coexist with the same mass and charge, which configuration has the higher entropy \cite{Dias:2021afz}?

\begin{acknowledgments}

The authors thank Funda\c{c}\~ao Amaz\^onia de Amparo a Estudos e Pesquisas (FAPESPA),  Conselho Nacional de Desenvolvimento Cient\'ifico e Tecnol\'ogico (CNPq) and Coordena\c{c}\~ao de Aperfei\c{c}oamento de Pessoal de N\'{\i}vel Superior (Capes) -- Finance Code 001, in Brazil, for partial financial support. M.P. and L.C. thank the University of Sheffield, in the United Kingdom, and University of Aveiro, in Portugal, respectively, for the kind hospitality during the completion of this work. This work has further been supported by the European Union's Horizon 2020 research and innovation (RISE) programme H2020-MSCA-RISE-2017 Grant No. FunFiCO-777740 and by the European Horizon Europe staff exchange (SE) programme HORIZON-MSCA-2021-SE-01 Grant No. NewFunFiCO-101086251. S.D.~acknowledges financial support from the Science and Technology Facilities Council (STFC) under Grant No.~ST/X000621/1 and Grant No.~ST/W006294/1.  

\end{acknowledgments}


\bibliographystyle{apsrev4-2}
\bibliography{abg_bound_states}

\end{document}